\documentclass[sn-mathphys,Numbered]{sn-jnl}

\usepackage{graphicx}%
\usepackage{multirow}%
\usepackage{amsmath,amssymb,amsfonts}%
\usepackage{amsthm}%
\usepackage{mathrsfs}%
\usepackage[title]{appendix}%
\usepackage{xcolor}%
\usepackage{textcomp}%
\usepackage{manyfoot}%
\usepackage{booktabs}%
\usepackage{algorithm}%
\usepackage{algorithmicx}%
\usepackage{algpseudocode}%
\usepackage{listings}%
\usepackage{colortbl}
\usepackage{tcolorbox}
\newcommand{\answer}[1]{
\vspace{2mm}\noindent\fbox{%
    \parbox{.97\columnwidth}{%
        {#1}
    }%
}\vspace{2mm}}

\newcommand{\nff}[1]{{\textcolor{blue}{#1}}}

\theoremstyle{thmstyleone}%
\theoremstyle{thmstyletwo}%

\theoremstyle{thmstylethree}%

\begin{document}

\title[Article Title]{An Extensive Replication Study of the ABLoTS Approach for Bug Localization}


\author*[1]{\fnm{Feifei} \sur{Niu}}\email{niufeifei@smail.nju.edu.cn}

\author[1]{\fnm{Enshuo} \sur{Zhang}} \email{2575357413@qq.com}

\author[2]{\fnm{Christoph} \sur{Mayr-Dorn}}\email{christoph.mayr-dorn@jku.at}

\author[3]{\fnm{Wesley} \sur{K. G. Assunção}}\email{wguezas@ncsu.edu}

\author[4]{\fnm{Liguo} \sur{Huang}}\email{lghuang@smu.edu}

\author[1]{\fnm{Jidong} \sur{Ge}}\email{gjd@nju.edu.cn}

\author[1]{\fnm{Bin} \sur{Luo}}\email{luobin@nju.edu.cn}

\author[2]{\fnm{Alexander} \sur{Egyed}}\email{alexander.egyed@jku.at}

\affil*[1]{\orgdiv{State Key Laboratory for Novel Software Technology}, \orgname{Nanjing University}, \orgaddress{\city{Nanjing}, \country{China}}}

\affil[2]{\orgdiv{Institute for Software Systems Engineering}, \orgname{Johannes Kepler University}, \orgaddress{\city{Linz}, \country{Austria}}}

\affil[3]{\orgdiv{Department of Computer Science}, \orgname{North Carolina State University}, \orgaddress{\city{North Carolina}, \country{USA}}}

\affil[4]{\orgdiv{Department of Computer Science}, \orgname{Southern Methodist University}, \orgaddress{\city{Dallas}, \state{Texas}, \country{USA}}}


\abstract{
Bug localization is the task of recommending source code locations (typically files) that contain the cause of a bug and hence need to be changed to fix the bug. Along these lines, information retrieval-based bug localization (IRBL) approaches have been adopted, which identify the most bug-prone files from the source code space. In current practice, a series of state-of-the-art IRBL techniques leverage the combination of different components (e.g., similar reports, version history, and code structure) to achieve better performance. ABLoTS is a recently proposed approach with the core component, TraceScore, that utilizes requirements and traceability information between different issue reports (i.e., feature requests and bug reports) to identify buggy source code snippets with promising results. To evaluate the accuracy of these results and obtain additional insights into the practical applicability of ABLoTS, we conducted a replication study of this approach with the original dataset and also on two extended datasets (i.e., additional Java dataset and Python dataset). \nff{The original dataset consists of 11 open source Java projects with 8,494 bug reports.}
The extended Java dataset includes 16 more projects comprising 25,893 bug reports and corresponding source code commits. The extended Python dataset consists of 12 projects with 1,289 bug reports. While we find that the TraceScore component\nff{, which is} the core of ABLoTS\nff{,} produces comparable or even better results with the extended datasets, we also find that \nff{we cannot reproduce the ABLoTS results, as reported in its original paper}, due to an overlooked side effect of incorrectly choosing a cut-off date that led to test data leaking into training data with significant effects on performance. Additionally, we conduct experiments to assess the performance of various composers \nff{that aggregate scores from different components, revealing} that Logistic Regression, fixed weight, and CombSUM outperform the other composers across all three datasets, while decision tree and random forest exhibited subpar performance.
}




\maketitle

\section{Introduction}\label{sec1}

A software bug refers to an error, fault, or flaw that produces unexpected results or causes a system to behave unexpectedly~\cite{tan2014bug}. A bug may cause the system to crash or become vulnerable to security attacks~\cite{aslan2017mitigating, piessens2002taxonomy}. Bugs are a common phenomenon. For example, a Mozilla triager complained that ``every day, almost 300 bugs appear that need triage''~\cite{anvik2005coping}. Considering the severe consequences and frequent occurrences, bugs need to be responded to promptly and coped seriously. To this end, various techniques to assist this process have been suggested, for example, defect prediction~\cite{kim2007predicting, zhang2009investigation}, bug triaging~\cite{anvik2006should, bettenburg2008duplicate}, bug localization~\cite{ciborowska2022fast, huo2019deep}, and bug fixing~\cite{li2019dfix, jeffrey2009bugfix}.

Bug localization, which \nff{refers to} identifying the parts of source code that cause the bug and need to be changed in order to fix it, is one of the main challenges when solving bugs in practice~\cite{loyola2018bug}. However, finding the buggy files from the source code can become a daunting task~\cite{wang2014version}, especially in large projects consisting of thousands of source code files. To help to deal with this issue, researchers proposed several automatic approaches for bug localization~\cite{lukins2010bug, zhou2012should, wang2014version, wang2016amalgam+}.

Among existing approaches for bug localization, there is a series of them that leverage bug reports for better localization~\cite{wang2014version, wang2016amalgam+, zhou2012should}, since bug reports often contain rich information that allows us to infer the bug's location. Approaches that utilize the textual content of bug reports are generally described as information retrieval-based bug localization (IRBL). For a given bug report, IRBL finds and ranks code snippets that may be relevant to the bug report~\cite{akbar2020large}, which is usually done by calculating the similarity between the bug report and source code~\cite{akbar2020large}. For example, Saha et al.~\cite{saha2013improving} propose the BLUiR approach that extracts structured information (e.g., class names, method names, variable names, and comments) from source code and calculates the textual similarity between the source code and bug reports to retrieve buggy files. However, there exists a lexical gap between bug reports and source code files~\cite{mcmillan2011exemplar}. The terms used to describe the bug in the bug report may not match the terms used in class names, methods names, variable names, or comments. Not surprisingly, textual similarity by itself will not necessarily yield good results~\cite{wang2014version}. 

To improve the performance of bug localization, state-of-the-art approaches leverage multiple sources of information. Wang et al. propose the AmaLgam approach, which combines code structure, similar bug reports, and version history~\cite{wang2014version}. Another approach, namely BRTracer+, leverages bug reports similarity and stack trace from bug reports for bug localization~\cite{wong2014boosting}. Youm et al. integrate stack trace information with all those pieces of information used by AmaLgam~\cite{youm2015bugblia}. Additionally, AmaLgam+ leverages five sources of information, namely version history, similar bug reports, code structure, stack trace, and reporter information~\cite{wang2016amalgam+}. 

Rath et al. presented a new approach, named ABLoTS, that leverages not only similar bug reports, version history, code structure, but also similar non-bug reports, like feature requests, enhancements, and tasks, as well as traceability information between bug reports and other types of issues~\cite{rath2018analyzing}. Rath et al. reused the structure of AmaLgam, but proposed TraceScore to replace the similar bug reports component, and additionally decided to use a decision tree (DT) for dynamically combining the recommendations from the individual components. The experimental evaluation showed that ABLoTS greatly outperforms AmaLgam.

Although the original study by Rath et al.~\cite{rath2018analyzing} showed encouraging results (with no other state-of-the-art approaches exhibiting better performance~\cite{bench4bl,li2022empirical}), there are no replications in the literature that confirm its outstanding performance. Additionally, there are no studies that investigate whether the performance also holds for other dataset\nff{s}, i.e., that evaluate the generalization of ABLoTS. 
A replication study is helpful and necessary to verify experimental results from previous studies~\cite{shepperd2018role}. They are a key aspect of empirical software engineering, as they bring evidence that observations made can hold (or not) under other conditions~\cite{carver2010towards}. Extensive and independent evaluations are also necessary to reach industrial adoption and practice~\cite{haben2021replication,da2014replication}.

In a previous work~\cite{niu2023ablots}, we present a literal and conceptual replication~\cite{gomez2014understanding} of the ABLoTS approach. We replicated the experiments as closely as possible to the initial procedures. We also run the experiment on \nff{a} new Java dataset without changing anything else, to see how well the results hold up.
Thus, we first re-implemented TraceScore, the core component of ABLoTS, and checked the replicability of the results on the original dataset. Then, we replicated the overall ABLoTS framework on the original dataset. Additionally, we investigated the TraceScore's and ABLoTS' generalizability on the new Java dataset including 16 more projects comprising 25,893 bug reports and corresponding source code commits. \nff{In general, our replication results show that TraceScore is replicable and generalizable under specific settings under a \textit{relaxed cut-off date} (i.e., use fixed date as cut-off date). 
However, ABLoTS is neither replicable on the original dataset nor on a larger dataset~\cite{rath2019seoss}. Specifically, we observed that the implementation of ABLoTS reused a subcomponent from prior work (AmaLgam~\cite{wang2014version}) that incorrectly sets a cut-off date, which leads to test data leaking into training data.}

In this paper, we report an extension of our previous work~\cite{niu2023ablots}. While our initial replication focused only on Java projects, in this extension, we introduce a Python dataset, and also other factors constant, to see how well the results hold up on new projects and new programming language.
The new dataset and new programming language are based on two extended datasets from SEOSS~33~\cite{rath2019seoss} and BuGL~\cite{muvva2020bugl}. Additionally to the new dataset, in this work, we also investigate the performance of various fusion methods for combining ABLoTS' three main scoring components.

The contributions of this paper are:

\begin{enumerate}
    \item An empirical investigation showing that the TraceScore component is replicable and generalizable, thus strengthening confidence that relations between bug reports and feature requests yield useful information for bug localization.
    \item A failed attempt to replicate the promising results of the ABLoTS approach, thereby showing that bug localization still needs significant research efforts and is not ready for practical application. Additionally, we present  the major reason why replication failed, thereby highlighting the challenge of reusing research results.
    \item Valuable findings for further studies on IRBL: 1) Combining different components can yield better results, however, researchers should always be careful about choosing the proper cut-off date. 2) AmaLgam's implementation of BugCache may not be as useful as expected, since they adopted a wrong cut-off date. 3) Preliminary exploration of the effectiveness of different composers reveals that Logistic Regression (LR), fixed weight and CombSUM perform well across the three datasets. In contrast, DT and Random Forest (RF) exhibit poorer performance.
    \item Replication package and experimental results\footnote{\url{https://github.com/feifeiniu-se/Replication2}} to replicate our experiment and evaluate the ABLoTS approach.
\end{enumerate}

Our work is organized according to standard replication report guidelines for software engineering studies~\cite{carver2010towards}. This is an external and independent replication study without any of the authors of the original paper taking part in the replication process.


The rest of this paper is organized as follows. Section~\ref{sec:originalstudy} summarizes the original study, approach, evaluation, and achieved results. Section~\ref{sec:methodology} elaborates our replication study design, research questions, and dataset. The experimental results are presented and discussed in Section~\ref{sec:discussion}. Section~\ref{sec:validity} discusses threats to validity. Related work is presented in Section~\ref{sec:relatedwork}, followed by the conclusion of this work in  Section~\ref{sec:conclusion}.


\section{Replicated Study}\label{sec:originalstudy}
\begin{figure}[!ht]
\setlength{\abovecaptionskip}{0.cm}
\centering{\includegraphics[width=0.7\linewidth]{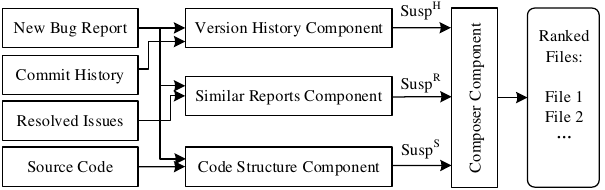}}
\caption{Components of ABLoTS.}
\label{fig:components}
\end{figure}

This work is a replication of the ABLoTS approach proposed by Rath et al.~\cite{rath2018analyzing}, which consists of four components shown in Fig.~\ref{fig:components}, namely \textbf{Version History Component}, \textbf{Similar Reports Component}, \textbf{Code Structure Component} and \textbf{Composer Component}. 
In this section, we provide an overview of the ABLoTS approach.
\nff{We firstly present the whole framework of the ABLoTS approach (Section~\ref{sec:ablots}), followed by the TraceScore component that is at the center of ABLoTS approach, encapsulated in the \textbf{Similar Reports Component} (described in Section~\ref{sec:tracescore}). Then, we present the utilized evaluation metrics (Section~\ref{sec:evaluation}) as well as the dataset as used in the original study (Section~\ref{sec:dataset}).} Finally, we summarize the reported experimental results (Section~\ref{sec:performance}). 
 
 \begin{figure}[!ht]
 \setlength{\abovecaptionskip}{0.cm}
\centering{\includegraphics[width=.5\linewidth]{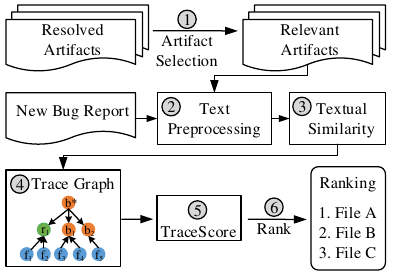}}
\caption{TraceScore Component.}
\label{fig:tracescore}
\end{figure}

\subsection{ABLoTS Approach}\label{sec:ablots}

As shown in Fig.~\ref{fig:components}, the overall ABLoTS approach consists of four components: 1) similar reports component, 2) version history component, 3) code structure component, and 4) composer component. Rath et al.~\cite{rath2018analyzing} implemented TraceScore as a similar reports component, and they reused the version history component, code structure component, and composer component without changes. They are briefly describe\nff{d} below.

\textbf{Version History Component} uses BugCache~\cite{kim2007predicting, rahman2011bugcache}, to predict which files are likely to be buggy in the future. BugCache takes commit history as input and outputs a list of files with a high ``suspiciousness'' score. To this end, it firstly identifies bug-fixing commits (commits whose commit messages contain the word ``fix'' or ``bug'') that were committed within $k$ days prior to the submission of the new bug report \textit{b*}. Then the suspiciousness score of each file $f$ is calculated by  Eq.~\ref{eq:bugcache}, where $f$ is one of the buggy files in commit $c \in C$, $t_c$ is the elapsed time in days between the commit \textit{c} and when the bug report was filed. \textit{k} was set to 15 (days) according to Wang et al.~\cite{wang2014version}.

\begin{equation}\label{eq:bugcache}
    Susp^H(f,b*) = \sum_{c\in C \wedge f\in c}^{}\frac{1}{1+e^{12(1-((k-t_c)/k))}}
\end{equation}

\nff{\textbf{Similar Reports Component} is based on the assumption that similar bugs will be caused by similar source code snippets. Hence, by identifying similar bugs reports and inspecting which files were changed in their bug-fixing commits, one can obtain a list of files indicating the bug location. ABLoTS approach implemented TraceScore as the similar reports component. Specifically, it introduces a calculation scheme for the similar reports component compared with SimiScore~\cite{zhou2012should}.}

\textbf{Code Structure Component} leverages BLUiR~\cite{saha2013improving} to identify files from source code space according to the similarity between source code files and bug report \textit{b*}. It outputs a ranked list of files with a suspiciousness score $Susp^S(f,b*)$.

\textbf{Composer Component} aggregates the three suspiciousness scores obtained by the first three components, i.e., $Susp^R$, $Susp^H$, $Susp^S$, and outputs the final results. Instead of adopting a fixed weight scheme for the three scores as done by Wang et al.~\cite{wang2014version, wang2016amalgam+}, ABLoTS applied Weka's~\cite{hall2009weka} J48 DT to learn the best combination. For training, the classification algorithm takes $Susp^R(f,b*)$, $Susp^H(f,b*)$, $Susp^S(f,b*)$ as the features, and whether that file $f$ was changed as part of the bug fix or not as the classification result. For each project separately, they trained the classifier on 80\% of the bug reports that were resolved and evaluated ABLoTS on the remaining 20\% that were resolved after the 80\% cut-off deadline. 

\subsection{TraceScore Component}\label{sec:tracescore}

TraceScore is one of the main components of the ABLoTS approach. 
It mainly consists of six steps, as shown in Fig.~\ref{fig:tracescore}. Given a new bug report \textit{b*}, TraceScore takes previously resolved issue reports (including bug reports \textit{B} and feature requests \textit{R}) as input. \textbf{Step 1} is artifact selection, based on two criteria, namely \textit{time domain} and\textit{ number of modified files}. For the \textit{time domain}, bug reports $b \in B$ and feature requests $r \in R$ that are fixed within ``one year before \textit{b*} was filed'' to ``the date when \textit{b*} was filed'', would be retained. As for \textit{number of modified files}, only bug reports $b \in B$ that modify no more than 10 Java files and feature requests $r \in R$ that modify no more than 20 Java files will be retained. The reasons for adoption of these two criteria and their validity are explained in Section 6 of the original study. 
\textbf{Step~2} utilizes commonly used preprocessing techniques to build a document-term-matrix~\cite{manning2008introduction} of the filtered artifacts from Step 1. Then,  in \textbf{Step 3}, TraceScore calculates the cosine similarity between \textit{b*} and each artifact. In \textbf{Step 4}, a trace graph is created, with \textit{b*} as the root node, linked to sub-graphs of different artifacts, by the edges indicating textual similarity between \textit{b*} and each artifact (if there is a trace link between \textit{b*} and artifact, the edge is set to 1). Each artifact traces further to the files that are part of a corresponding commit in the version control system. In this way, b* is indirectly linked to a potentially large set of source code files, that need subsequent ranking, where the ranking happens on the basis of a \textbf{TraceScore} between each file and \textit{b*} which is calculated by Eq.~\ref{eq:tracescore} in \textbf{Step~5}. Finally, \textbf{Step~6} sorts all the source code files linked to \textit{b*} according to TraceScore and outputs the ranked list. A higher score indicates a higher likelihood of that file being relevant.

\begin{equation}\label{eq:tracescore}
    Susp^R(s,b*) = \sum_{a_i \in \left \{ {a|s \in fix(a)} \right \}} \frac{sim(a_i,b*)^2}{|fix(a_i)|}  
\end{equation}

\subsection{Evaluation Metrics}\label{sec:evaluation}

To evaluate the effectiveness of ABLoTS, Rath et al. adopted three metrics: 

\textbf{Top \textit{k}}~\cite{schutze2008introduction} measures the percentage of bug reports in which at least one of the buggy files is in top \textit{k} ranked files, where \textit{k}$=$1, 5, 10. 

\textbf{Mean Average Precision (MAP)}~\cite{manning2008introduction} is calculated as the mean of the Average Precision over all queries. Average Precision of a given bug report aggregates precision of positively recommended files as:

\begin{equation}
AP = \sum_{i=1}^{N}\frac{P(i) * pos(i)}{\# \, of \, positive \, instances}   
\end{equation}

where $i$ is a rank of the ranked files, $N$ is the number of ranked files and $pos(i)$ $\in$ \{0,1\} indicates whether the \textit{i}th  file is a buggy file or not. $P(i)$ is the precision at a given top $i$ files:

\begin{equation}
    P(i) = \frac{\# \, of \, buggy \, files \, in \, top \, i}{i}
\end{equation}

\textbf{Mean Reciprocal Rank (MRR)}~\cite{voorhees1999trec} computes the average of the reciprocal of the positions of the first correctly located buggy file in the ranked files, following this equation:

\begin{equation}
MRR = \frac{1}{Q}\sum_{|Q|}^{q=1}\frac{1}{rank_i} 
\label{eq:mrr}
\end{equation}

\subsection{Original Dataset}\label{sec:dataset}
In the original study, Rath et al. contributed a dataset~\cite{rath2018replication} consisting of 15 open source Java projects with 13,581 bug reports and 9,219 feature requests. Firstly, they collected issues reports (i.e., bug reports and feature requests), as well as the dependency trace links from Jira~\cite{jira2018}, and downloaded source code of these projects from GitHub~\cite{git2018}. Then the heuristic proposed in~\cite{bachmann2009software} was applied to create links between issues and commits. The ABLoTS approach was evaluated based on this dataset.

\subsection{Achieved Performance Originally Reported}\label{sec:performance}
The \nff{reported} performance by ABLoTS is shown in Table~\ref{tab:performance}, which is the average on 15 projects. The average MAP and MRR of TraceScore is 20.2\% and 26\%, while that of ABLoTS is 48.8\% and 54.5\%, respectively. Specifically, ABLoTS exhibits the ability to accurately identify at least one buggy file for 48.7\% of the bug reports when considering only the top-ranked files. Moreover, in the top 5 ranked files, ABLoTS successfully identifies at least one buggy file for 64.9\% of the bug reports. Rath et al. revealed that TraceScore benefits from leveraging non-bug issues as well as traceability information. It can outperform two state-of-the-art similar reports based approaches: SimiScore~\cite{zhou2012should} and CollabScore~\cite{ye2014learning}. The overall ABLoTS framework leverages DT as composer and outperforms the AmaLgam framework~\cite{wang2014version} which leverages fixed weight composer.

\begin{table}[htbp]
\setlength{\abovecaptionskip}{0.cm}
\caption{Original, reported results~\cite{rath2018analyzing}}
\label{tab:performance}
\begin{tabular}{lccccc}
\hline
 Algorithm & MAP   & MRR   & Top 1 & Top 5 & Top 10 \\ \hline
TraceScore & 0.202 & 0.260 & 0.174 & 0.350 & 0.436  \\
ABLoTS     & 0.488 & 0.545 & 0.487 & 0.610 & 0.649  \\ \hline
\end{tabular}
\end{table}

\section{Replication Methodology}\label{sec:methodology}

The goal of this replication study is to investigate whether the results based on the TraceScore component and ABLoTS approach are replicable and generalizable to different projects and languages. Furthermore, we explore the performance of different composers on aggregating the three suspiciousness scores. To this end, we 1) replicate the component and the approach on a subset of the original dataset; 2) apply the component and the approach on 16 more Java projects and 12 more Python projects; and 3) evaluate the performance of different composers by fusing the three components in diverse ways. We provide a description of the extended datasets in Section~\ref{sec:newdata}. Subsequently, we outline our research questions in Section~\ref{sec:rqs}, followed by detailed explanations of the replication methodology (Section~\ref{sec:rq1} through Section~\ref{sec:rq3}). We reuse the same evaluation metrics as Rath et al., including MAP, MRR, and Top 1, 3, and 5 (Section~\ref{sec:evaluation}).

This study is considered to be an external~\cite{carver2010towards} replication study of the original study, since none of the authors took part in the replication process. However, we reused 11 projects of the original dataset to verify the results.\footnote{The other four projects from the original dataset were excluded due to some missing commits on GitHub.}

\subsection{Dataset}\label{sec:newdata}

For the replicability validation, we reuse the dataset provided for replication by Rath et al.~\cite{rath2018replication}, which we refer to as ``Original-Java''. However, by the time we carried out the replication study, many commits from four projects (i.e., Axis2, Hadoop, Infinispan, and Pig) were no longer available on GitHub, and neither \nff{were} part of the original replication package. Hence, as we could not obtain \nff{the} complete commit history for BugCache, we excluded these four projects from the analysis in this paper. 

As stated in Section~\ref{sec:rqs}, the generalizability evaluation comes from two aspects: 1) additional dataset of the same language and 2) additional dataset of different programming language. To this end, we picked two additional datasets: 1) SEOSS 33 dataset~\cite{rath2019seoss} and 2) BuGL dataset~\cite{muvva2020bugl}. The SEOSS 33 dataset includes \nff{18 additional} projects and 36,482 bug reports out of which we could not use 2 projects due to the same issue of non-accessible commits. We refer to this dataset as ``Extended-Java''. This extended dataset also includes the 15 original projects from the replication package~\cite{rath2018replication}. We choose this dataset because it not only links bug reports to commit code change, but also includes traceability information between bug reports and non-bug issues, which caters to our needs perfectly. The BuGL dataset is a large-scale cross-language dataset for bug localization. BuGL consists of more than 10,000 bug reports drawn from open source projects written in four programming languages, namely C, C++, Java, and Python. In this research, we only select Python projects in BuGL dataset, which includes 12 open source projects with 1,289 bug reports. We refer to this dataset as ``Extended-Python''. 

The reason for choosing Python for our study is that it is one of the most popular programming languages~\cite{b2016learning, benton2019defexts}, even outperformed Java according to the 2023 Stack Overflow \nff{D}eveloper \nff{S}urvey. However, there is no feature requests or traceability information between issues contained in the Python dataset. Thus, we can only leverage the similar bug reports for TraceScore, which may bring bias to our results.
Details about the extended dataset are shown in Table~\ref{tab:dataset}, which shows that the Java projects tend to be larger compared to Python projects when considering the number of  bug reports and source code files.

Apart from the information in the dataset, we additionally collected version information for each project. In each commit, developers may modify a file, add a new file, or remove an old file. 
Removed files are obsolete and should not appear in the recommendation of a new bug report. However, the similar reports component leverages historical issues, which may be pointing to no longer existing files. In this way, they may bias the prediction results. To address this issue, we determine for each commit which files exist just prior to this commit. Files in this set can only be used as the candidates to recommend the bug's location. 

\begin{table}[htbp]
\caption{Characteristics of the Extended-Java and Extended-Python Dataset.}
\label{tab:dataset}
\begin{tabular}{lccc||lcc}
\hline\hline
\begin{tabular}[c]{@{}c@{}} Java \\ Projects\end{tabular}    & \begin{tabular}[c]{@{}c@{}}\# Bug \\ Reports\end{tabular} & \begin{tabular}[c]{@{}c@{}}\# Non-bug \\ Reports\end{tabular} & \# Files & \begin{tabular}[c]{@{}c@{}} Python \\ Projects\end{tabular} & \begin{tabular}[c]{@{}c@{}}\# Bug \\ Reports\end{tabular} & \# Files \\  \hline
ARCHIVA     & 371            & 411               & 1024  &    ERTBOT                 & 170  & 345 \\
CASSANDRA   & 3571           & 2813              & 3595 &    COMPOSE                 & 155  & 82  \\
ERRAI       & 267            & 194               & 3703  &    DJANGO\_R.\_F. & 153  & 155 \\
FLINK       & 1350           & 2351              & 13192 &    FLASK                   & 53   & 73  \\
GROOVY      & 1933           & 1017              & 1375 &    KERAS                   & 51   & 198 \\
HBASE       & 4581           & 5171              & 4657 &    MITMPROXY               & 107  & 371 \\
HIBERNATE   & 1947           & 1706              & 11720  &    PIPENV                  & 136  & 857 \\
HIVE        & 4776           & 4326              & 4657 &    REQUESTS                & 75   & 35  \\
JBOSS-T.-M. & 331            & 489               & 4204  &    SCIKIT-LEARN            & 1082 & 767 \\
KAFKA       & 639           & 1149              & 3423   &    SCRAPY                  & 112  & 304 \\
LUCENE      & 3773           & 5324              & 5482 &    SPACY                   & 55   & 643 \\
MAVEN       & 760            & 574               & 1381 &    TORNADO                 & 40   & 114 \\
RESTEASY    & 345            & 228               & 3866  &&&     \\
SPARK       & 328           & 7022              & 1055  &&&    \\
SWITCHYARD  & 451            & 759               & 2953  &&&     \\
ZOOKEEPER   & 470            & 471               & 900  &&&     \\ \hline \hline
\end{tabular}
\end{table}

\subsection{Research Questions}\label{sec:rqs}
The replication study aims mainly at answering the following research questions (RQ):
\begin{itemize}
    \item RQ1. How effective is TraceScore in identifying bug-relevant source code files?
    \begin{itemize}
        \item RQ1.1 Are we able to replicate the original performance of the TraceScore component?
        \item RQ1.2 Does the TraceScore component yield similar performance when applied to additional Java projects?
        \item RQ1.3 How does TraceScore perform on Python projects?
    \end{itemize}
    
    \item RQ2. How effective is ABLoTS for bug localization?
    \begin{itemize}
        \item RQ2.1 Are we able to replicate the results of the ABLoTS approach?
        \item RQ2.2 Does the ABLoTS approach yield similar performance when applied to additional Java projects?
        \item RQ2.3 How does ABLoTS perform on Python projects?
    \end{itemize}

    \item RQ3. How do different composers perform when combining three components?
\end{itemize}

In our study, RQ1 and RQ2 are adapted from those addressed in the original study, which involve the main contribution of the Rath et al. study~\cite{rath2018analyzing}. Specifically, RQ1 is adapted from the first question of the original study, which evaluates the TraceScore component. RQ2 is adapted from fourth question of the original study, which evaluates the ABLoTS approach. For each research question, we \nff{analyze} two dimensions, namely 1) replicability and 2) generalizability. The core difference of the dimensions is the dataset being used for evaluation. For the replicability validation, we evaluate on the same projects with the original study, to see if our replication results are consistent with the original results. As for the generalizability validation, it can be divided into two levels. The first level focuses on the same programming language, namely Java, but on \nff{16 additional projects, to see if the approach is applicable to other projects as well}. The second level extends to different programming language, particularly focusing on \nff{12 additional} Python projects, \nff{to see how the approach performs on different programming languages}. 
The other two RQs in the original study (i.e., second and third questions) mainly investigate the effectiveness of artifacts selection, which is irrelevant of our goal, so we do not include them in this study. 

Our RQ3 aims at investigating the performance of different composers at combining the three suspiciousness scores, including fixed weight, Multi-layer Perception (MLP), DT, RF, LR, CombSUM, CombMNZ, CombANZ, CorrB\nff{,} and Borda Count. Aside from merely aiming to improve the composition mechanism, the reason for studying the composer performance emerges from our finding in RQ2 that the original ABLoTS composer was trained on leaking data (i.e., the wrong cut-off date for BugCache) and hence is expected to no longer work effectively when data leakage does not occur. We are thus interested whether an improved composer may achieve the original ABLoTS performance or whether the fixed weights from the AmaLgam approach yield the best possible scoring result.

\subsection{Procedure to answer RQ1}\label{sec:rq1} 
This research question mainly focuses on the main contribution of Rath et al., i.e., TraceScore for the similar reports component.
Since the original source code is not available, we followed the procedures proposed in the original paper (illustrated in Section~\ref{sec:tracescore}) as close as possible to duplicate all facets of TraceScore. 
Specifically, for each project, we sort all the issues according to the resolved date. Then we split all the bug reports 80:20, with the latter 20\% used as the test set to recommend buggy files. 

Similar to the original study, we filter the number of related bug reports and features as well as commits based on age and size from which to obtain a recommendation.
For each bug report $b*$ in the test set, we consider only bug reports (and features) $b$ that occurred before $b*$ as determined by the following condition: $b.fixed\_date > b*.created\_date - 365 \, days$. However, we are of the opinion that there is another constraint that also should be satisfied:  $b.fixed\_date < b*.created\_date$, which means that only bug reports fixed before $b*$  were filed should be retained. These two settings describe the following two recommendation situations: the former describes the bug localization mechanism called shortly before fixing the bug, close to the bug report's closing date, while the latter describes a recommendation immediately made upon bug creation. For our replication, we were unable to determine whether the authors only adopted the first constraint (denoted as \textit{relaxed cut-off date}) or adopted both constraints (denoted as \textit{strict cut-off date}). We conducted the replication with both relaxed and strict cut-off date to understand the impact the additional constraint has on the results.  
Then, we select bug reports/features requests according to the \textit{number of modified files} identified in their commits. We exclude issues that modify more than 10 files for bug reports and more than 20 files for non-bug reports. We then build up the trace graph from these issue subset as shown in Fig.~\ref{fig:tracescore} in Step~4. The edges between the root node $b*$ and other artifacts are calculated using cosine similarity~\cite{baeza1999modern}. When an issue explicitly links to another issue, then the link weight overrides the cosine similarity and becomes 1. 
With the trace graph, the TraceScore between each file node $s$ and $b*$, $TraceScore(s,b*)$ is calculated. Finally, all the files according to their tracescore, we will get the ranked list for $b*$. 

Then we evaluate our replication on the two extended datasets. For the extended datasets, we apply preprocessing (Step 2 in Section~\ref{sec:tracescore}) to be consistent with the original dataset and to fit the replication. Specifically, for each issue, we preprocess the text including both summary and description (Step 2 in Section~\ref{sec:tracescore}). We utilize NLTK library~\cite{nltk} in Python for preprocessing, including stop words \nff{removal}, camel case splitting, lower casing\nff{,} and stemming. Then the preprocessed texts are converted into TF-IDF~\cite{jones1972statistical} vector with the sklearn library~\cite{sklearn}. For the source code, we exclude non-source code files based on the file name extension and only retain source code files (``*.java'' for Extended-Java and ``*.py'' for Extended-Python). For each file changed in each commit, the extended dataset contains the old name and the new name for this file. \nff{Following} the original study, we only utilize the new name for each file, which means removed files will be excluded for each commit. 

\subsection{Procedure to answer RQ2}\label{sec:3-c}

The ABLoTS approach is essentially an ensemble of three components, namely similar reports, version history, and code structure, as shown in Fig.~\ref{fig:components}. 
As described in the original study, the ABLoTS approach is an evolved version of AmaLgam~\cite{wang2014version} with two main differences: 1) it replaces the similar reports component with TraceScore; and 2) it applies a dynamic suspiciousness score combination (instead of the former static one). At the time of conducting the replication, there is no available implementation for the whole framework. We, therefore, replicated the framework along the following lines. 

\textbf{Version History.} As mentioned in Section~\ref{sec:ablots}, the version history component is implemented by BugCache, which is proposed by Kim et al.~\cite{kim2007predicting}. BugCache maintains the modification history of files to predict buggy-prone files in the future. It proved that more recently and frequently modified files are more likely to be buggy in the future. Rahman et al. proposed a simpler version of BugCache~\cite{rahman2011bugcache}, which only maintains a short history of file modification. Google's developers adapted Rahman et al.'s algorithm on their large systems~\cite{lewis2013does, timmurphy.org}. AmaLgam adapted Google's well-tested algorithm with a version history component. We reused AmaLgam's implementation of BugCache,\footnote{\url{https://sites.google.com/view/mambalab/projects/amalgam}} but made the following modifications:

\begin{enumerate}
    \item The BugCache version used in AmaLgam was written in Java, while we manually translated it to Python to be compatible with our implementation.
    \item In their paper, Wang et al.~\cite{wang2014version} \nff{explained} that the approach identifies commits that are committed 15 days before \textit{the new bug report is created}. However, after checking the source code, we found that the implementation utilized the bug report's \textit{resolved date} as the cut-off date to obtain previously committed commits within 15 days. We contacted the authors, and they agreed that the bug report's \textit{creation dates} should have been adopted. Therefore, in our implementation, we used \textit{the creation date} for all our experiments.
    \item To identify bug-fixing commits, Wang et al. proposed that commit logs should match regular expression regex: \textit{(.*fix.*)\textbar(.*bug.*)}. Considering that some programming languages (e.g., Java and Python) are case-sensitive, we firstly convert commit logs into lowercase, which is missing in the original implementation. What is more, according to our observation of the dataset, some bug-fixing commit logs may not contain keywords like ``fix'' or ``bug''. However, they might start with the bug report's ID. To this end, we also include commits that start with any bug ID in their logs, to identify bug-fixing commits more accurately. AmaLgam's authors also agree with us on this. This adapted selection of commits only affects the commits used for BugCache, but not any other component in ABLoTS.
\end{enumerate}

\textbf{Code Structure.} Code Structure metrics are obtained with BLUiR~\cite{saha2013improving}, which calculates the similarity between a new bug report and the code structure of a source code file. It takes the summary and description of a bug report as two separate parts and extracts class names, method names, variable names, and comments of a source code file represented as an Abstract Syntax Tree (AST). Then it indexes and searches buggy files based on the Indri toolkit~\cite{strohman2005indri}. In this paper, instead of replicating our own BLUiR tool, we used the implementation\footnote{\url{https://github.com/exatoa/bench4bl}} from an empirical study by Lee et al.~\cite{bench4bl} to obtain the $Susp^S(s, b*)$ score. Lee et al. implementation were originally designed for Java language. Since our experiments also involve Python files, we parsed Python files with the AST module to extract the code structure. Then we used the Indri\footnote{\url{https://sourceforge.net/projects/lemur/files/lemur/}} tool to calculate the similarity between bug reports and code structure of Python files.

\textbf{Composer.} ABLoTS applied the J48 DT with default pruning settings to classify source code files for bug reports. Specifically, for each $b*$, there are multiple candidate source code files $s$ for recommendation. For each (s,b*), there will be a label $C$ $\in$ \{$true$, $false$\} indicating whether the file $s$ is modified to fix $b$ or not. For training, the classifier takes the $Susp^R(s,b*)$, $Susp^H(s, b*)$, and $Susp^S(s,b*)$ scores for each (s, b*) as feature and $C$ as \nff{the} label. For test data, instead of outputting a label indicating true or false, the probability of $s$ being \textit{true} (i.e., $s$ is modified by $b*$) is utilized. Then for each $b*$, all the files are ranked according to the probability score.

\nff{For each project}, Rath et al. sorted all the bug reports by resolved date and took the first 80\% bug reports as training data, and the remaining 20\% as test data. To mitigate the influence of imbalanced training data, ABLoTS used Weka's sub-sampling to under-sampling the training data.

Since our replication is based on Python, we chose the popular open source Python library sklearn~\cite{sklearn} for the DT classifier, and \textit{RandomUnderSampler} in the Imblearn library~\cite{imblearn} for under-sampling. Essentially they are the same algorithm with the original study, but just implemented by different libraries. We assume that this will not cause significant difference to the result as we used exactly the same training data as in the original paper (i.e., rather than sampling our own set of training data we utilized the precalculated suspiciousness scores and classification result from the replication package to obtain a trained DT).

We applied the same procedure on the original dataset and the extended datasets.
After completing the replication of the entire framework on the original dataset, we assess the performance of ABLoTS on the extended \nff{Java and Python} datasets to evaluate how it performs.

\subsection{Procedure to answer RQ3}\label{sec:rq3}

Rath et al.~\cite{rath2018analyzing} chose DT as the composer and claimed that DT outperformed fixed weight utilized by AmaLgam~\cite{wang2014version, wang2016amalgam+}. The essence of ABLoTS lies in adopting an aggregation strategy, where the three different scores: $Susp^R$, $Susp^H$, $Susp^S$ for each $<$$b*$, $s$$>$ pair are aggregated to obtain the final relevance score. Since there also exist other supervised and unsupervised aggregation strategies, in this study, we would like to explore the performance of different strategies, including unsupervised: fixed weight, CombSUM~\cite{fox1993combining,fox1994combination}, CombMNZ~\cite{fox1993combining,fox1994combination}, CombANZ~\cite{fox1993combining,fox1994combination}, CorrB~\cite{wu12} and Borda count~\cite{AslamM01}, and supervised: MLP, DT, RF and LR. CombSUM, CombMNZ, CombANZ, and Borda count are also rank fusion methods that have been investigated for fusing fault localizers~\cite{lucia2014fusion}. Different methods may be more suitable for different scenarios. Thus, it is important to choose the most appropriate aggregation approach. 

\textbf{Fixed Weight} has been proved to be effective for IRBL~\cite{wang2016amalgam+, wang2014version, youm2015bugblia}. The suspiciousness score for the source code file $s$ is calculated according to Eq.~\ref{eq:susp1} and Eq.~\ref{eq:susp2}, where the value of $a$ and $b$ are set to 0.2 and 0.3 as per prior work.

\begin{equation}\label{eq:susp1}
 Susp^{R,S}(s) = a*Susp^{R}(s) + (1-a)*Susp^{S}(s)
\end{equation}

\begin{equation}\label{eq:susp2}
\begin{split}
 & Susp^{R,S,H}(s) = b*Susp^{H}(s) + \\
 & (1-b)*Susp^{R,S}(s)
\end{split}
\end{equation}

\textbf{CombSUM}~\cite{fox1993combining,fox1994combination} is a simple rank fusion method where the scores of documents from different lists are summed, and the documents are ranked based on the total sum. While straightforward, CombSUM assumes that all methods contribute equally, which may not always be the case.

\textbf{CombANZ} is another rank fusion method that combines different scores by computing the average of the non-zero scores.

\textbf{CombMNZ}~\cite{fox1993combining, fox1994combination} is a variant of CombANZ. It involves multiplying the summation of scores for a given element by the number of techniques that assign a non-zero score to that element.

\textbf{CorrB}~\cite{wu12} is a correlation-based method that calculates the weight of a technique by assessing the overlap of its list of the top-N most suspicious program elements with lists generated by other techniques.

\textbf{Borda Count}~\cite{AslamM01} is a popular rank fusion method. In Borda count, each document in a ranked list receives a score equal to the sum of the positions it holds in the individual lists. The document with the highest Borda score is ranked first in the integrated list.

Apart from the above explained unsupervised rank fusion methods, there are also supervised methods that have been used to learn feature importance for classification problems. In this study, we use the commonly used classification algorithms: MLP, RF and LR. For each bug report and source code file pair $<$$b*$,$s$$>$, if $s$ is related to $b*$, then the ground truth is set to be 1, otherwise 0. The optimization function of classification algorithms is to learn if $s$ is related to $b*$.

\section{Results and Discussion}\label{sec:discussion}

We present below the results of our replication study, organized for answering each RQ. Then, we discuss the overall findings and implication or our work.

\subsection{RQ1. How effective is TraceScore in identifying bug-relevant source code files?}

\textbf{RQ1.1: Replicability.} 
We carried out the replication according to Section~\ref{sec:rq1}. Results on the original dataset are as shown in Table~\ref{tab:rq1.1-relax}.
The performance impact of using the \textit{strict cut-off date} is on average around 17\% lower than using the \textit{relaxed cut-off date}.

\begin{table}[htbp]
\setlength{\abovecaptionskip}{0.cm}
\caption{TraceScore performance on the original dataset.}
\label{tab:rq1.1-relax}
\begin{tabular}{lccccc}
\hline \hline
\multicolumn{6}{c}{Relaxed Cut-off Date}          \\ \hline 
PROJECTS & MAP   & MRR   & Top 1 & Top 5 & Top 10 \\ \hline
DERBY    & 0.124 & 0.240 & 0.149 & 0.340 & 0.404  \\
DROOLS   & 0.183 & 0.383 & 0.276 & 0.502 & 0.615  \\
HORNETQ  & 0.134 & 0.241 & 0.130 & 0.352 & 0.481  \\
IZPACK   & 0.170 & 0.229 & 0.156 & 0.328 & 0.422  \\
KEYCLOAK & 0.125 & 0.234 & 0.152 & 0.323 & 0.418  \\
LOG4J2   & 0.182 & 0.271 & 0.191 & 0.360 & 0.416  \\
RAILO    & 0.138 & 0.202 & 0.117 & 0.267 & 0.350  \\
SEAM2    & 0.134 & 0.195 & 0.141 & 0.244 & 0.288  \\
TEIID    & 0.194 & 0.278 & 0.188 & 0.385 & 0.465  \\
WELD     & 0.102 & 0.208 & 0.098 & 0.312 & 0.420  \\
WILDFLY  & 0.108 & 0.185 & 0.116 & 0.268 & 0.326  \\ \hline
Average  & 0.145 & 0.242 & 0.156 & 0.335 & 0.419  \\ \hline \hline
\multicolumn{6}{c}{Strict Cut-off Date}           \\ \hline 
PROJECTS & MAP   & MRR   & Top 1 & Top 5 & Top 10 \\ \hline
DERBY    & 0.084 & 0.158 & 0.096 & 0.219 & 0.272  \\
DROOLS   & 0.171 & 0.37  & 0.265 & 0.467 & 0.603  \\
HORNETQ  & 0.105 & 0.207 & 0.093 & 0.315 & 0.444  \\
IZPACK   & 0.101 & 0.152 & 0.094 & 0.219 & 0.297  \\
KEYCLOAK & 0.081 & 0.16  & 0.082 & 0.241 & 0.323  \\
LOG4J2   & 0.165 & 0.256 & 0.18  & 0.315 & 0.382  \\
RAILO    & 0.131 & 0.194 & 0.117 & 0.25  & 0.35   \\
SEAM2    & 0.099 & 0.159 & 0.103 & 0.212 & 0.263  \\
TEIID    & 0.140 & 0.222 & 0.135 & 0.331 & 0.412  \\
WELD     & 0.103 & 0.201 & 0.098 & 0.304 & 0.411  \\
WILDFLY  & 0.085 & 0.146 & 0.087 & 0.217 & 0.268  \\
Average  & 0.115 & 0.202 & 0.123 & 0.281 & 0.366  \\ \hline \hline
\end{tabular}
\end{table}

To find out which implementation most likely was adopted by the original implementation, we performed a pairwise t-test on the 11 projects, comparing both replication results against the reported results in~\cite{rath2018analyzing} to establish statistically whether these results can be considered to be the same. According to the pairwise t-test, the \textit{relaxed cut-off date} is closer to the original implementation. The pairwise t-test results presented in Table~\ref{tab:ttest} show that for MRR, Top1, Top5, and Top10, there is no significant difference while for MAP. Thus, we have to reject the null hypothesis for the \textit{relaxed cut-off date}: the average MAP reported by Rath et al. is 32\% higher than our replication result. For the remaining four evaluation metrics, there is no significant difference, \nff{with} the mean values \nff{being} statistically the same. So we conclude that with the \textit{relaxed cut-off date} TraceScore can be considered replicable, while with the \textit{strict cut-off date} it cannot be considered replicable, since we cannot achieve statistically comparable or better results. 

To give benefit to doubt, we adopted the \textit{relaxed cut-off date} for the remainder of the replication and generalization investigations. However, in practice, the choice between \textit{relaxed cut-off date} and \textit{strict cut-off date} is artificial as only commits available at the time the bug localization mechanism is applied are considered for producing the recommendation. 



\begin{table}[htbp]
\setlength{\abovecaptionskip}{0.cm}
\caption{Pairwise t-test between relax constraint result and original result.}
\label{tab:ttest}
\begin{tabular}{ccccc}
\hline
\multirow{2}{*}{Metrics} & \multicolumn{2}{c}{Pairs} & \multirow{2}{*}{Deviation} & \multirow{2}{*}{P value} \\ 
                         & Original   & Replication  &                            &                          \\ \hline
MAP                      & 0.191      & 0.145        & 0.05                       & 0.000**                  \\
MRR                      & 0.248      & 0.242        & 0.01                       & 0.522                    \\
Top 1                    & 0.163      & 0.156        & 0.01                       & 0.407                    \\
Top 5                    & 0.336      & 0.335        & 0.00                       & 0.924                    \\
Top 10                   & 0.419      & 0.419        & 0.00                       & 0.99                     \\ \hline
\multicolumn{4}{l}{*p \textless 0.05 **p \textless 0.01}\\
\end{tabular}
\end{table}

\textbf{RQ1.2 \& RQ1.3: Generalizability.}
The evaluation results based on the extended Java and Python dataset are shown in Table~\ref{tab:rq1.2}. The average MAP, MRR, Top 1, Top 5 and Top 10 are 18.3\%, 28.4\%, 19.6\%, 38.4\%, 47.3\% on Java dataset, and 25.6\%, 34.6\%, 23.5\%, 48.4\%, 59.2\% on Python dataset, respectively. The average results are around 20\% (Top 1) $\sim$ 40\% (MAP) higher on Python projects than on Java projects. The MAP values ranges from 4.4\% to 32.5\% in the Java dataset, and from 5.2\% to 50.5\% in the Python dataset. The MRR values stretches from 11.4\% to 47.3\% for the Java dataset, and lies between 11.4\% and 56.9\% for the Python dataset.

\begin{table}[htbp]
\setlength{\abovecaptionskip}{0.cm}
\caption{TraceScore performance on the extended datasets.}
\label{tab:rq1.2}
\begin{tabular}{lccccc}
\hline \hline
\multicolumn{6}{c}{Extended-Java dataset} \\ \hline
PROJECTS    & MAP   & MRR   & Top 1  & Top 5  & Top 10 \\ \hline
ARCHIVA     & 0.134 & 0.22  & 0.147 & 0.28  & 0.413 \\
CASSANDRA   & 0.218 & 0.333 & 0.222 & 0.453 & 0.551 \\
ERRAI       & 0.059 & 0.15  & 0.093 & 0.204 & 0.296 \\
FLINK       & 0.18  & 0.305 & 0.207 & 0.415 & 0.522 \\
GROOVY      & 0.325 & 0.393 & 0.271 & 0.522 & 0.625 \\
HBASE       & 0.236 & 0.352 & 0.25  & 0.455 & 0.561 \\
HIBERNATE   & 0.118 & 0.231 & 0.172 & 0.3   & 0.359 \\
HIVE        & 0.264 & 0.38  & 0.267 & 0.506 & 0.599 \\
JBOSS-T.-M. & 0.136 & 0.247 & 0.164 & 0.373 & 0.433 \\
KAFKA       & 0.296 & 0.473 & 0.367 & 0.578 & 0.688 \\
LUCENE      & 0.201 & 0.32  & 0.228 & 0.419 & 0.494 \\
MAVEN       & 0.162 & 0.222 & 0.132 & 0.316 & 0.382 \\
RESTEASY    & 0.101 & 0.202 & 0.101 & 0.348 & 0.435 \\
SPARK       & 0.31  & 0.383 & 0.273 & 0.545 & 0.576 \\
SWITCHYARD  & 0.044 & 0.114 & 0.088 & 0.121 & 0.198 \\
ZOOKEEPER   & 0.149 & 0.226 & 0.149 & 0.309 & 0.436 \\ \hline
\textbf{Average}     & \textbf{0.183} & \textbf{0.284} & \textbf{0.196} & \textbf{0.384} & \textbf{0.473} \\ \hline \hline
 \multicolumn{6}{c}{Extended-Python dataset} \\
\hline
PROJECTS    & MAP   & MRR   & Top 1  & Top 5  & Top 10 \\ \hline
CERTBOT                 & 0.264   & 0.369       & 0.235   & 0.500       & 0.706       \\
COMPOSE                 & 0.325   & 0.423       & 0.290   & 0.548       & 0.677       \\
DJANGO\_R.\_F. & 0.505   & 0.555       & 0.484   & 0.613       & 0.710       \\
FLASK                    & 0.279   & 0.365       & 0.182   & 0.636       & 0.727       \\
KERAS                    & 0.149   & 0.171       & 0.091   & 0.273       & 0.455       \\
MITMPROXY               & 0.133   & 0.247       & 0.182   & 0.364       & 0.409       \\
PIPENV                  & 0.310   & 0.569       & 0.464   & 0.714       & 0.786       \\
REQUESTS                 & 0.328   & 0.343       & 0.200   & 0.600       & 0.600       \\
SCIKIT-LEARN           & 0.305   & 0.373       & 0.290   & 0.465       & 0.530       \\
SCRAPY                  & 0.190   & 0.297       & 0.217   & 0.304       & 0.522       \\
SPACY                    & 0.235   & 0.321       & 0.182   & 0.545       & 0.727       \\
TORNADO                  & 0.052   & 0.114       & 0.000   & 0.250       & 0.250       \\ \hline
\textbf{Average}  & \textbf{0.256} & \textbf{0.346} & \textbf{0.235} & \textbf{0.484} & \textbf{0.592}\\ \hline \hline
\end{tabular}
\end{table}

\begin{figure*}[!ht]
\setlength{\abovecaptionskip}{0.cm}
\centering{\includegraphics[width=0.7\linewidth]{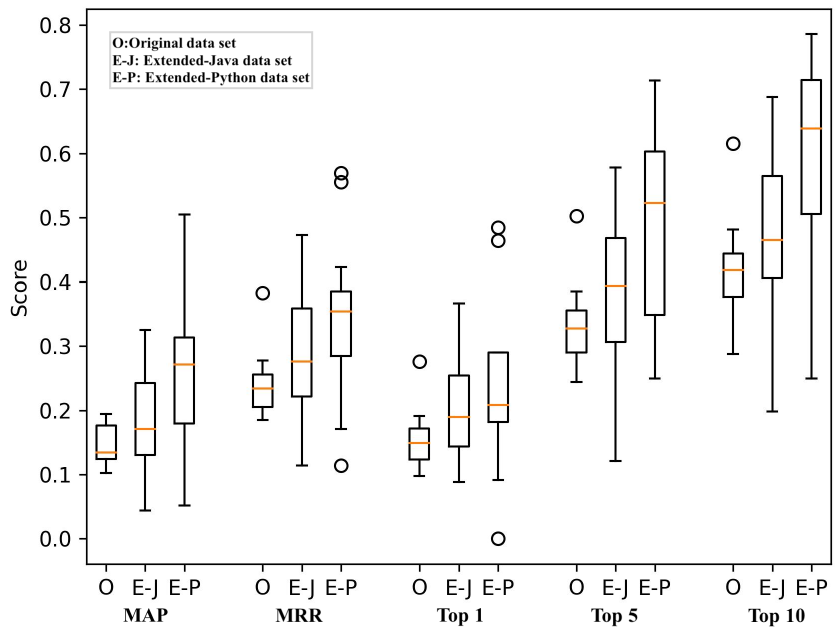}}
\caption{Box plots of TraceScore on original and extended Java and Python datasets.}
\label{fig:box-tracescore}
\end{figure*}

In order to confirm if there is a difference between the distribution of the original results and extended results, we leverage the two-sample Kolmogorov-Smirnov test (K-S test)~\cite{massey1951kolmogorov}, which is used to test whether two samples come from the same underlying one-dimensional probability distribution. For each evaluation metric, we perform a two-sample K-S test, with one sample being the results from the original dataset and the other sample being the results from one of the two extended datasets. Results  are shown in the ``TraceScore-J'' and ``TraceScore-P'' columns of Table~\ref{tab:ks1} for the Extended-Java dataset and the Extended-Python dataset, individually. On the Extended-Java dataset, all the p-values are greater than 5\%, indicating that the two samples come from the same distribution. For the Extended-Python dataset, however, all the p-values are less than 5\%, thus we cannot assume that the two samples come from the same distribution.
The box plot in Fig.~\ref{fig:box-tracescore} shows the data value on all five metrics, from which we can see that on the Extended-Java dataset, TraceScore yields slightly higher median and wider variations, while on the Extended-Python dataset, TraceScore yields much higher median, maximum, and minimum, which indicates TraceScore yields higher performance on the Extended-Python dataset than on the original Java dataset. The average over the Extended-Java dataset is about 12\% (Top 10) $\sim$ 27\% (MAP) higher than on the original dataset, while that of the Extended-Python dataset is about 41\% (Top 10) $\sim$ 77\% (MAP) higher than on the original dataset.

\begin{figure*}[!ht]
\setlength{\abovecaptionskip}{0.cm}
\centering{\includegraphics[width=0.7\linewidth]{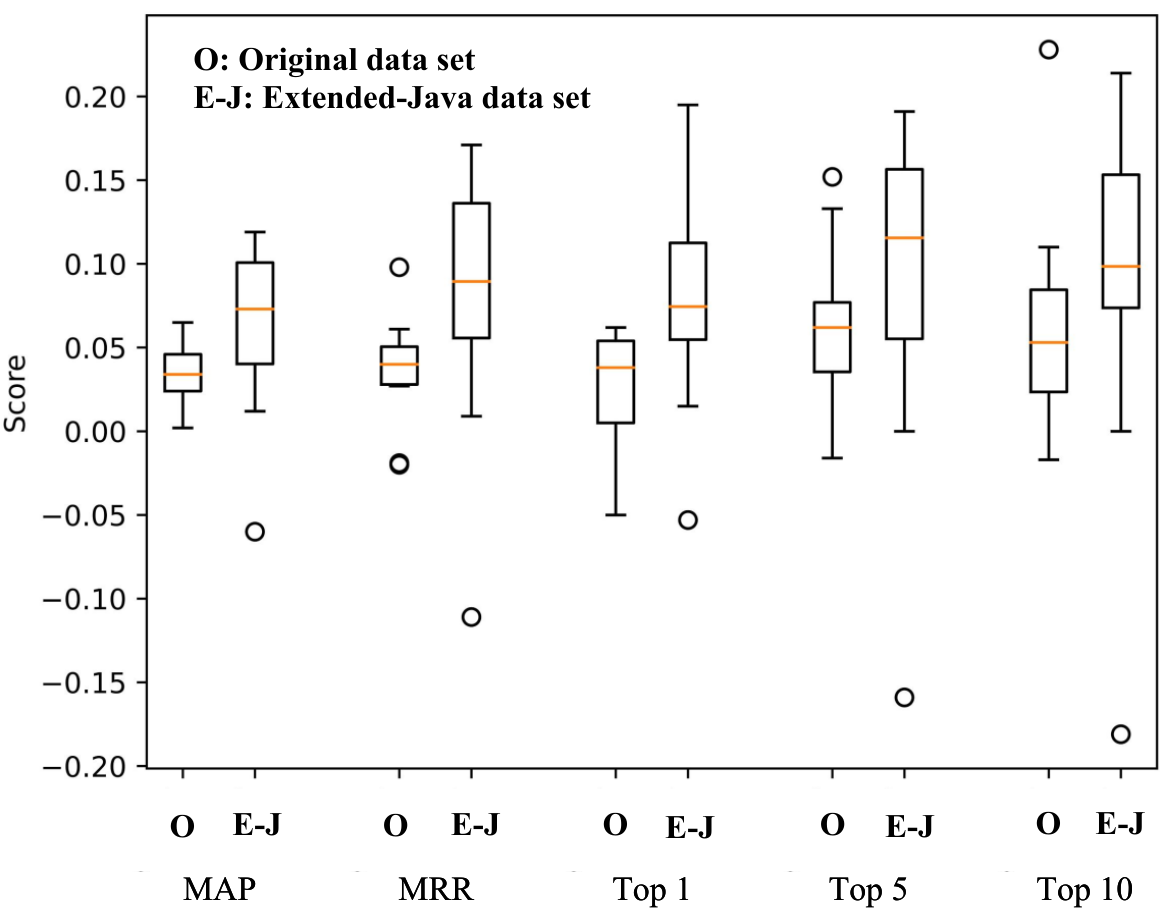}}
\caption{Box plots illustrating the improvement of TraceScore over SimiScore on both the original and extended Java datasets.}
\label{fig:box-improvement}
\end{figure*}

We also investigate the improvement of TraceScore over the same baseline as in the original paper.\footnote{Since the main improvement of TraceScore over SimiScore lies in the leveraging of requirements and traceability information, but there is no such information in Extended-Python dataset, we do not compare the improvement over the Extended-Python dataset} With SimiScore~\cite{zhou2012should} as baseline, we obtain the improvement of TraceScore over SimiScore on both original and Extended-Java dataset. The ``Improvement'' column of Table~\ref{tab:ks1} shows the results of the  K-S test. Given the p-values, the improvement of MRR, Top 1, and Top 10 on the original dataset and the extended dataset are very likely to come from different distributions. To this end, from the box plot in Fig.~\ref{fig:box-improvement} for improvement, we can observe a much higher median, maximum, and minimum, which indicates TraceScore yields higher performance improvement on the Extended-Java dataset. 

We can therefore conclude that the performance of TraceScore also holds for a larger Java dataset or for another Python dataset, and we gain confidence that TraceScore's performance is generally achievable.

\begin{table}[htbp]
\setlength{\abovecaptionskip}{0.cm}
\caption{K-S test result.}
\label{tab:ks1}
\begin{tabular}{cccccccccc}
\hline
\hline
        &  & \multicolumn{2}{c}{TraceScore-Java} &  & \multicolumn{2}{l}{TraceScore-Python} &  & \multicolumn{2}{c}{Improvement} \\
Metrics &  & K-S test        & P value        &  & K-S test        & P value        &  & K-S test        & P value       \\ \hline
MAP     &  & 0.438           & 0.124          &  & 0.667           & 0.004          &  & 0.500           & 0.054         \\
MRR     &  & 0.409           & 0.175          &  & 0.659           & 0.006          &  & 0.693           & 0.002         \\
Top 1   &  & 0.409           & 0.175          &  & 0.561           & 0.039          &  & 0.625           & 0.007         \\
Top 5   &  & 0.409           & 0.175          &  & 0.576           & 0.023          &  & 0.443           & 0.115         \\
Top 10  &  & 0.415           & 0.159          &  & 0.659           & 0.006          &  & 0.540           & 0.028         \\ \hline \hline
\end{tabular}
\end{table}

\answer{
\textit{\textbf{Answering RQ1:} Under the relax cut-off date constraint, TraceScore is replicable and also can be generalized to both Java and Python extended datasets. However, under the strict cut-off date constraint, we cannot claim replicability as the performance is significantly lower than reported. }}

\subsection{RQ2. How effective is ABLoTS for bug localization?}

\textbf{RQ2.1: Replicability.}
ABLoTS's performance results on the original dataset are shown in Table~\ref{tab:rq2.1}. Compared to the results reported in the original paper (cf. Table~\ref{tab:performance}) we observe that our replication produces far worse results. MAP and MRR  are below 10\% for most projects. ABLoTS, which combines three scores, namely $Susp^R$, $Susp^H$, and $Susp^S$, does not even achieve the same results as the single $Susp^R$ score. This counterintuitive result motivated us to investigate in more detail how this outcome can be explained. 

For the strict replication, we trained the DT on the intermediate three scores (i.e., $Susp^R$, $Susp^H$, $Susp^S$) made available by Rath et al. in their replication package, For comparison, we also trained a separate DT from our own sample of files, their suspiciousness scores, and bug reports. Note that the original replication package just provided tuples of suspiciousness scores and classification results, but not which bug report and which files were used to obtain those suspiciousness scores. We, however, applied the same sampling criteria.

We inspected the original DT (i.e., the one obtained from the replication data) to obtain the average
feature importance (non-normalized) of each component: 0.037 for BLUiR, 0.377 for BugCache, and 0.018 for TraceScore. This indicates that BugCache almost exclusively determines the final classification result. In contrast, in the AmaLgam approach, which was used as a baseline for ABLoTS, the authors empirically set fixed weights for the three suspiciousness scores, which are 0.56 for BLUiR, 0.3 for BugCache and 0.14 for TraceScore. Our DT trained from scratch exhibited the following (non-normalized) feature importance: 0.243 for BLUiR, 0.007 for BugCache, 0.037 for TraceScore, which still does not yield as good results (see Table~\ref{tab:rq2.1}) as the fixed weights determined for AmaLgam.

\begin{figure}[!tp]
\setlength{\abovecaptionskip}{0.cm}
\centering{\includegraphics[width=\linewidth]{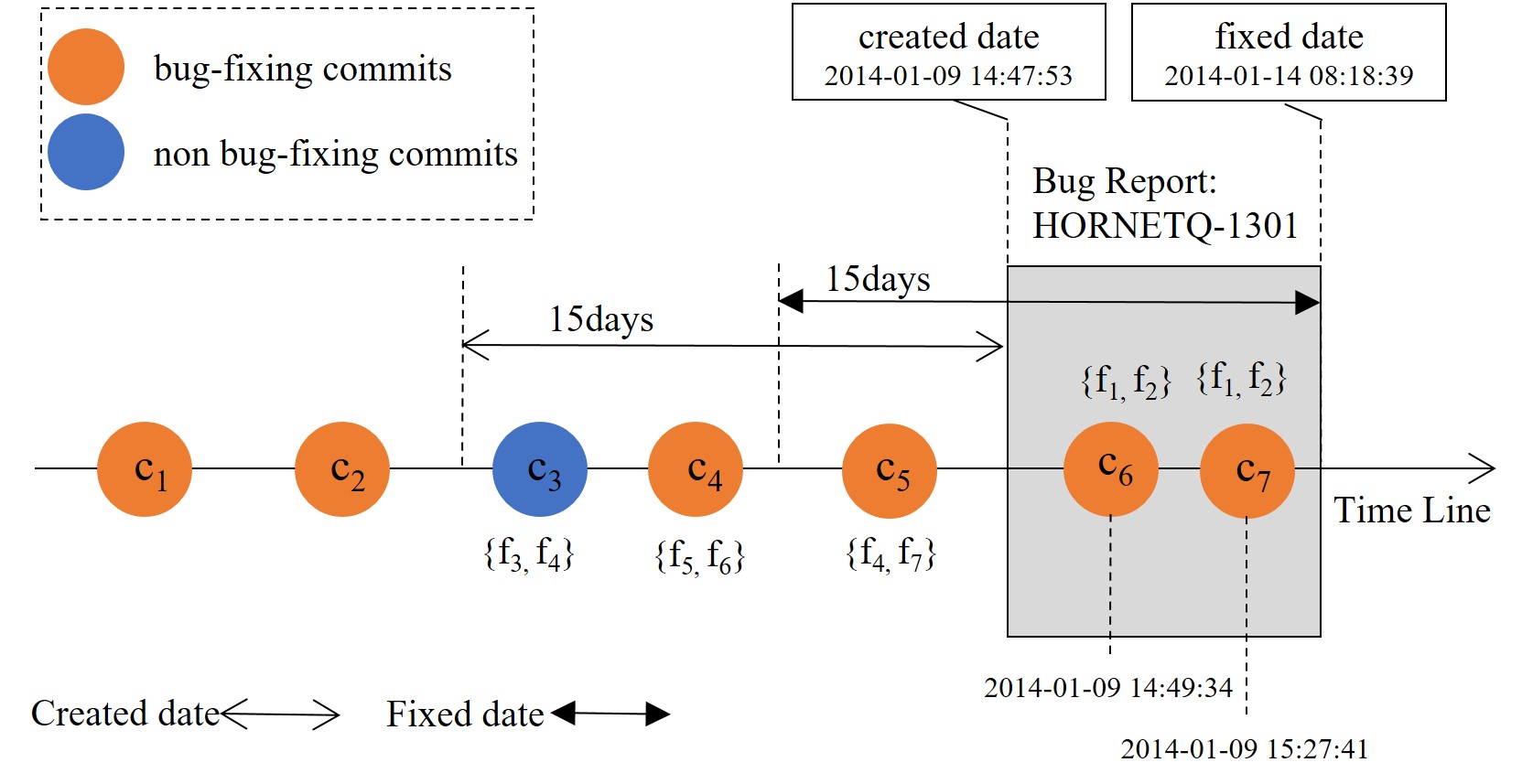}}
\caption{BugCache using created date VS using fixed date.}
\label{fig:bugcache}
\end{figure}

This discrepancy in feature importance values helped us identify the root cause for the difference in performance results.
Rath et al. adopted the implementation of BugCache by Wang et al.~\cite{wang2014version}, where the bug report's fixed date was utilized for the cut-off date, as shown in Fig.~\ref{fig:bugcache}. If one or more bug-fixing commits occurred within 15 days prior to the fixed date, BugCache would recommend the files within these commits (i.e., potentially exactly those files that were changed to fix the bug). However, in a realistic bug localization situation, any file recommendation would only be useful before any of those commits. Thus, for correct evaluation, these commits must not be used.

Fig.~\ref{fig:bugcache} illustrates such a situation. There is a bug report ``HORNETQ-1301'' created on 2014-01-09, and fixed on 2014-01-14. Two commits $c_6$ and $c_7$ were committed to fix this bug between the created date and fixed date, on 2014-01-09. When BugCache adopts the fixed date as the cut-off date and identifies bug-fixing commits within 15 days, then $c_4$, $c_5$, $c_6$, and $c_7$ would be taken into consideration and result in a high $Susp^H$ score, according to Eq.~\ref{eq:bugcache}. Doing so, the DT would learn that the scores by BugCache are very indicative of the actual classification result and hence assign it a high feature importance. However, in practice, $c_6$ and $c_7$ are unknown for predicting bug report ``HORNETQ-1301'', they are foreknowledge about the bug. The right way of implementing BugCache is using the creation date, or any date before the bug's first partial fix implementation. After contacting the authors of both ABLoTS and AmaLgam, AmaLgam's authors stated that they agreed with our finding and that they adopted the wrong date, while authors of ABLoTS stated that they directly reused AmaLgam's implementation.

The incorrectly derived $Susp^H$ scores thus greatly boost the result of the DT. When we utilized BugCache in the correct manner (i.e., use the created date as the cut-off date), DT did not yield results even close to the original performance (even when applying hyperparameter tuning). For comparison, we adopted AmaLgam's composer with a fixed weight for each component: 0.56 for BLUiR, 0.3 for BugCache, and 0.14 for TraceScore. The results of the fixed weight composer are shown in Table~\ref{tab:fix-old}, the average MAP, MRR, Top 1. Top 5, and Top 10 are 29.8\%, 43.3\%, 32\%, 56.3\% and 64\%, respectively. Compared to TraceScore, the results have been improved by 105.8\%, 78.7\%, 105.2\%, 68.4\%, and 52.9\%, respectively.

Aside from the DT feature importance values, a second discrepancy emerged when we investigated the evaluation dataset. In the replication package, the intermediary suspiciousness scores were provided not only as a training set for the DT but also as an evaluation set (i.e., the remaining 20\%). When we trained and evaluated with these two datasets, we could replicate the results. However, as outlined above, when obtaining the suspiciousness scores ourselves, we could not. The discrepancy we found was that the evaluation dataset contained far fewer evaluation data points (i.e., suspiciousness scores with their classification ground truth) than these projects contained source code files. In other words, for a particular bug, not all source code files were utilized for evaluation but just a subset.
Across all projects, the number of candidates ranges from 60 to 70, regardless of actual number of files in the respective project.
For the project HORNETQ, for example, even when we select only files for which a TraceScore suspiciousness score and a BLUiR suspiciousness score exist, we obtain around 4500 file candidates.
In addition, for some of these files the evaluation dataset does not provide any of the three suspiciousness scores at all, just the classification result. Hence, we could not establish how these file candidates have been filtered and why only a subset has been chosen. The paper does not describe this aspect, but rather refers to the evaluation design of Amalgam. 

In summary, we found that ABLoTS adopted the wrong cut-off date for BugCache due to having reused the component and configuration from AmaLgam without further investigation, resulting in the incorrect $Susp^H$ scores. 
Hence, we conclude that ABLoTS performance cannot be replicated.

\begin{table}[htbp]
\setlength{\abovecaptionskip}{0.cm}
\caption{ABLoTS performance on original dataset.}
\label{tab:rq2.1}
\begin{tabular}{lccccc}
\hline
PROJECTS    & MAP   & MRR   & Top 1  & Top 5  & Top 10 \\ \hline
DERBY    & 0.076 & 0.111 & 0.02  & 0.171 & 0.326 \\
DROOLS   & 0.049 & 0.06  & 0.023 & 0.054 & 0.097 \\
HORNETQ  & 0.057 & 0.067 & 0     & 0.056 & 0.185 \\
IZPACK   & 0.086 & 0.11  & 0.016 & 0.172 & 0.375 \\
KEYCLOAK & 0.029 & 0.05  & 0.006 & 0.044 & 0.101 \\
LOG4J2   & 0.065 & 0.072 & 0.011 & 0.067 & 0.146 \\
RAILO    & 0.06  & 0.077 & 0     & 0.1   & 0.283 \\
SEAM2    & 0.08  & 0.105 & 0.019 & 0.179 & 0.333 \\
TEIID    & 0.056 & 0.079 & 0.015 & 0.104 & 0.231 \\
WELD     & 0.02  & 0.024 & 0     & 0.018 & 0.027 \\
WILDFLY  & 0.03  & 0.04  & 0.007 & 0.036 & 0.087 \\ \hline
\textbf{Average}  & \textbf{0.055} & \textbf{0.072} & \textbf{0.011} &\textbf{ 0.091} & \textbf{0.199} \\ \hline
\end{tabular}
\end{table}

\begin{table}[htbp]
\setlength{\abovecaptionskip}{0.cm}
\caption{Fixed weight composer on original dataset.}
\label{tab:fix-old}
\begin{tabular}{lccccc}
\hline
PROJECTS    & MAP   & MRR   & Top 1  & Top 5  & Top 10 \\ \hline
DERBY    & 0.312 & 0.478 & 0.36  & 0.615 & 0.725 \\
DROOLS   & 0.272 & 0.464 & 0.339 & 0.607 & 0.712 \\
HORNETQ  & 0.37  & 0.555 & 0.426 & 0.704 & 0.778 \\
IZPACK   & 0.37  & 0.493 & 0.391 & 0.594 & 0.672 \\
KEYCLOAK & 0.234 & 0.377 & 0.247 & 0.525 & 0.595 \\
LOG4J2   & 0.391 & 0.541 & 0.416 & 0.719 & 0.753 \\
RAILO    & 0.286 & 0.398 & 0.267 & 0.567 & 0.65  \\
SEAM2    & 0.339 & 0.402 & 0.308 & 0.532 & 0.583 \\
TEIID    & 0.12  & 0.169 & 0.1   & 0.208 & 0.296 \\
WELD     & 0.252 & 0.445 & 0.33  & 0.562 & 0.634 \\
WILDFLY  & 0.334 & 0.441 & 0.333 & 0.565 & 0.645 \\ \hline
\textbf{Average}  & \textbf{0.298} & \textbf{0.433} & \textbf{0.320} & \textbf{0.563} & \textbf{0.640} \\ \hline
\end{tabular}
\end{table}

\textbf{RQ2.2 \& RQ2.3: Generalizability.} Since \nff{the evaluation results of the original paper presenting} ABLoTS is not replicable, exploring its performance on the extended dataset for generalizability evaluation would yield little insight. However, in order to explore how TraceScore would perform when jointly used with the other two components, like in AmaLgam~\cite{wang2014version, wang2016amalgam+}, we applied a fixed weight to aggregate the three scores. That is, the suspiciousness score for the source code file $s$ is calculated according to Eq.~\ref{eq:susp2}, where the value of $a$ and $b$ are set to 0.2 and 0.3 as per prior work.

\begin{figure*}[!ht]
\setlength{\abovecaptionskip}{0.cm}
\centering{\includegraphics[width=0.7\linewidth]{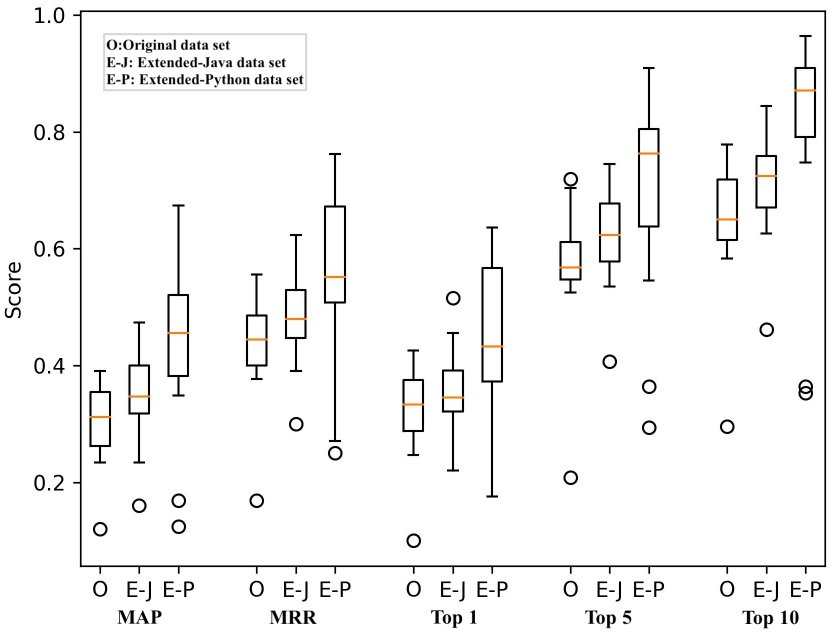}}
\caption{Box plots of fixed weight on original and extended Java and Python datasets.}
\label{fig:fixedweight}
\end{figure*}

The results of fixed weight are shown in Table~\ref{tab:fix-new}. On the additional 16 Java projects, the fixed weight composer can achieve an average MAP, MRR, Top 1, Top 5, Top 10 as 34.4\%, 47.7\%, 35.6\%, 62.1\% and 71.4\%, which improves over the single TraceScore by 87.8\%, 67.8\%, 81.8\%, 61.6\% and 50.8\%, respectively. Compared to the results on the original dataset, the average evaluation results over the extended dataset are 10\% $\sim$ 16\% higher (e.g., the average MAP is 34.4 vs 20.2). On the additional Python projects, the average MAP, MRR, Top 1, Top 5, and Top 10 are 43.5\%, 55.1\%, 43.9\%, 68.8\% and 78.5\%, respectively, which are around 10\% (Top 10) $\sim$ 26\% (MAP) higher than the average of additional Java projects. K-S test (Table~\ref{tab:ks2}) shows that on the extended Java dataset, all the p-values are greater than 5\%, so we should reject the hypothesis that the two samples come from different distributions. On the extended Python dataset, all the p-values are less than 5\%, thus the hypothesis that the two samples come from different distributions should be assumed to be true.
According to the box plot in Fig.~\ref{fig:fixedweight}, we can see that on the extended Java dataset, the distribution \nff{of} each metric is more concentrated, more similar, and the mean values are closer. On the extended Python dataset, the distribution \nff{of} each metric is more dispersed, with wider ranges, but the mean is significantly higher than that of the other two Java datasets.

Based on the results and analysis, we can conclude that the performance of the fixed weight composer also holds for a larger dataset, and we gain confidence in its generalizability.

\begin{table}[]
\setlength{\abovecaptionskip}{0.cm}
\caption{Fixed weight composer performance on the extended datasets.}
\label{tab:fix-new}
\begin{tabular}{lccccc}
\hline \hline
\multicolumn{6}{c}{Extended-Java dataset} \\ \hline
PROJECTS    & MAP   & MRR   & Top 1  & Top 5  & Top 10 \\ \hline
ARCHIVA     & 0.322 & 0.477 & 0.347 & 0.587 & 0.667 \\
CASSANDRA   & 0.335 & 0.462 & 0.330 & 0.622 & 0.741 \\
ERRAI       & 0.310 & 0.505 & 0.389 & 0.630 & 0.722 \\
FLINK       & 0.416 & 0.560 & 0.456 & 0.670 & 0.752 \\
GROOVY      & 0.388 & 0.458 & 0.331 & 0.618 & 0.726 \\
HBASE       & 0.398 & 0.528 & 0.398 & 0.697 & 0.778 \\
HIBERNATE   & 0.234 & 0.400 & 0.290 & 0.551 & 0.626 \\
HIVE        & 0.357 & 0.483 & 0.343 & 0.647 & 0.746 \\
JBOSS-T.-M. & 0.370 & 0.536 & 0.403 & 0.701 & 0.791 \\
KAFKA       & 0.474 & 0.623 & 0.516 & 0.742 & 0.844 \\
LUCENE      & 0.321 & 0.466 & 0.336 & 0.624 & 0.710 \\
MAVEN       & 0.337 & 0.416 & 0.296 & 0.546 & 0.671 \\
RESTEASY    & 0.257 & 0.391 & 0.275 & 0.536 & 0.638 \\
SPARK       & 0.406 & 0.496 & 0.379 & 0.606 & 0.712 \\
SWITCHYARD  & 0.160 & 0.300 & 0.220 & 0.407 & 0.462 \\
ZOOKEEPER   & 0.422 & 0.537 & 0.383 & 0.745 & 0.830 \\ \hline
\textbf{AVERAGE}     & \textbf{0.344} & \textbf{0.477} & \textbf{0.356} & \textbf{0.621} & \textbf{0.714} \\\hline \hline
\multicolumn{6}{c}{Extended-Python dataset} \\ \hline
PROJECTS    & MAP   & MRR   & Top 1  & Top 5  & Top 10 \\ \hline
CERTBOT        & 0.124       & 0.25        & 0.176       & 0.294       & 0.353       \\
COMPOSE        & 0.349       & 0.465       & 0.29        & 0.677       & 0.806       \\
DJANGO\_R.\_F. & 0.578       & 0.637       & 0.548       & 0.742       & 0.871       \\
FLASK          & 0.674       & 0.762       & 0.636       & 0.909       & 0.909       \\
KERAS          & 0.488       & 0.523       & 0.455       & 0.545       & 0.818       \\
MITMPROXY      & 0.43        & 0.551       & 0.409       & 0.818       & 0.909       \\
PIPENV         & 0.393       & 0.687       & 0.571       & 0.786       & 0.964       \\
REQUESTS       & 0.502       & 0.551       & 0.4         & 0.8         & 0.933       \\
SCIKIT-LEARN   & 0.422       & 0.533       & 0.41        & 0.668       & 0.747       \\
SCRAPY         & 0.481       & 0.667       & 0.565       & 0.783       & 0.87        \\
SPACY          & 0.169       & 0.271       & 0.182       & 0.364       & 0.364       \\
TORNADO        & 0.606       & 0.714       & 0.625       & 0.875       & 0.875       \\
\textbf{Average}  & \textbf{0.435} & \textbf{0.551} & \textbf{0.439} & \textbf{0.688} & \textbf{0.785} \\ \hline \hline
\end{tabular}
\end{table}

\begin{table}[htbp]
\setlength{\abovecaptionskip}{0.cm}
\caption{K-S test result of Fixed Weight.}
\label{tab:ks2}
\begin{tabular}{ccccccc}
\hline
\hline
&  & \multicolumn{2}{c}{Fixed Weight-Java}                      &  & \multicolumn{2}{l}{Fixed Weight-Python}                    \\
Metrics &  & \multicolumn{1}{c}{K-S test} & \multicolumn{1}{c}{P value} &  & \multicolumn{1}{c}{K-S test} & \multicolumn{1}{c}{P value} \\ \hline
MAP     &  & 0.313                        & 0.452                       &  & 0.750                        & 0.001                       \\
MRR     &  & 0.295                        & 0.512                       &  & 0.568                        & 0.031                       \\
Top 1   &  & 0.210                        & 0.856                       &  & 0.568                        & 0.031                       \\
Top 5   &  & 0.443                        & 0.115                       &  & 0.583                        & 0.017                       \\
Top 10  &  & 0.358                        & 0.289                       &  & 0.750                        & 0.001                       \\ \hline \hline

\end{tabular}
\end{table}

\answer{
\textit{\textbf{Answering RQ2:} The reported results of ABLoTS are not replicable, because of the incorrect use of the cut-off date in the BugCache component and the sub-optimal configuration of the composer. Consequently, we did not check the generalizability of ABLoTS, since applying an incorrect technique would provide little useful insight. However, with a fixed weight scheme, the results are generalizable on the extended dataset.}} 

\subsection{RQ3. How do different composers perform when combining three components?}
\begin{figure}[!ht]
\centering{\includegraphics[width=\linewidth]{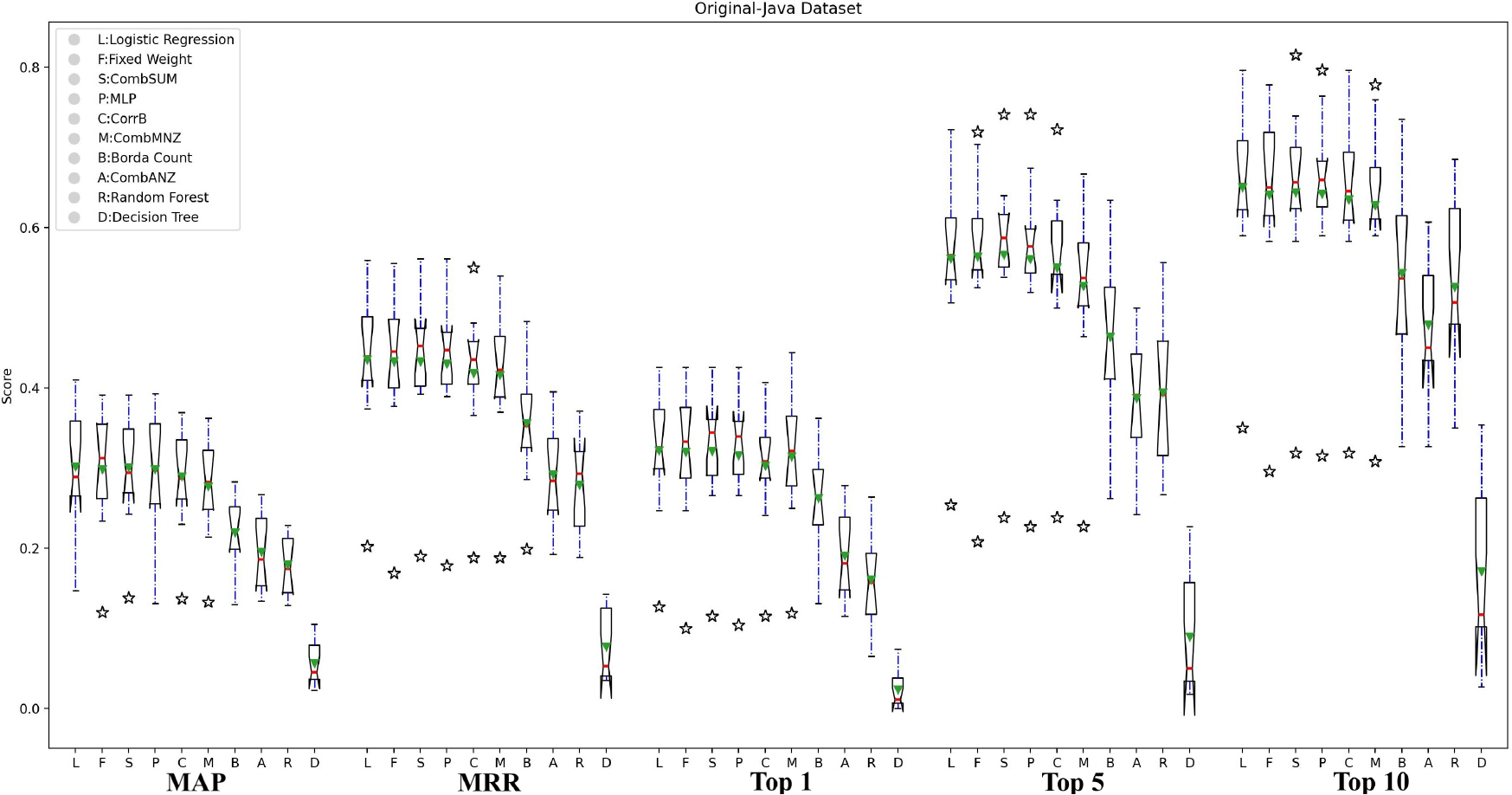}}
\caption{Performance of Different Aggregation Methods on Original Java Dataset.}
\label{fig:original-java}
\end{figure}

\begin{figure}[!ht]
\centering{\includegraphics[width=\linewidth]{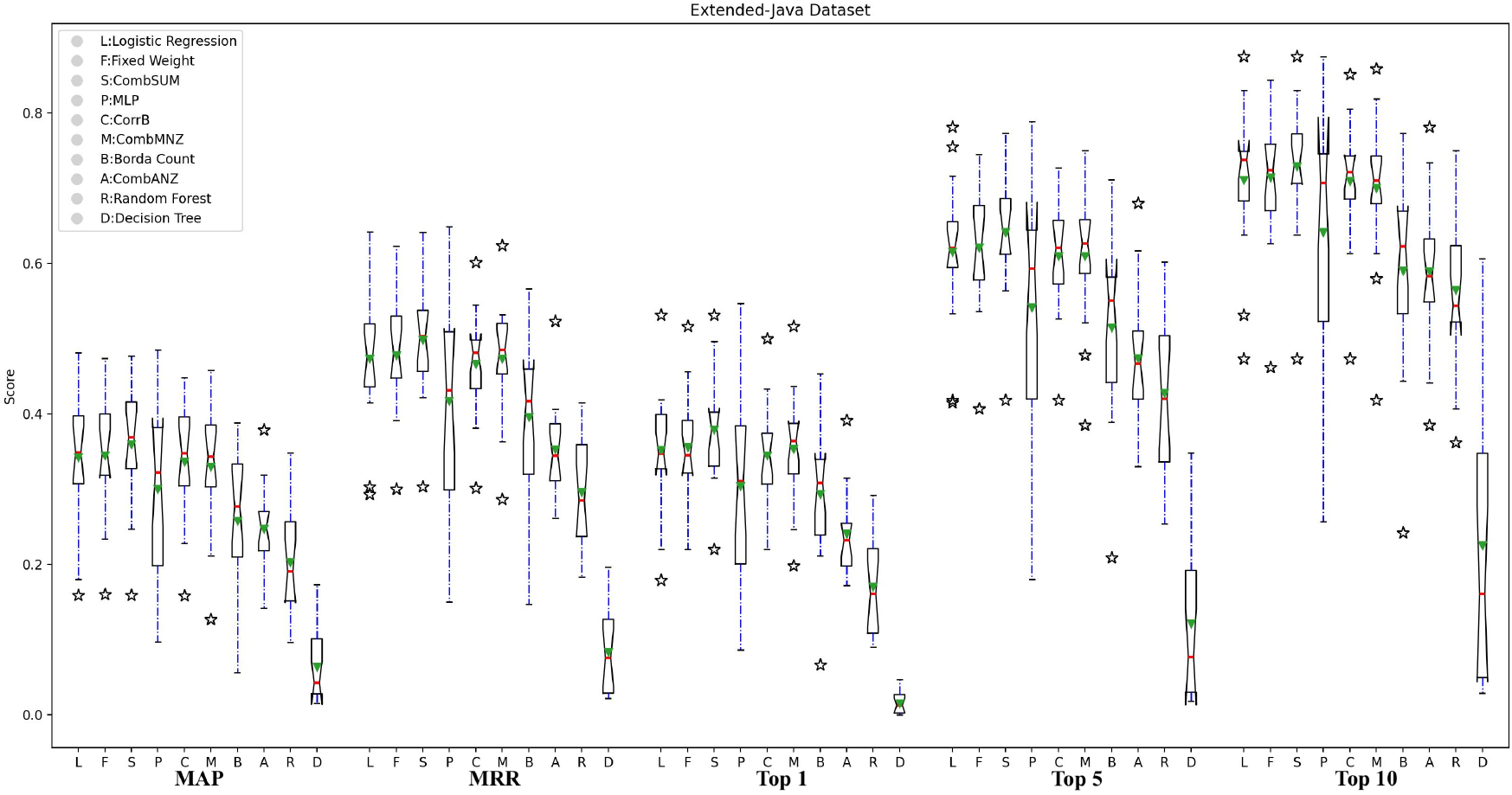}}
\caption{Performance of Different Aggregation Methods on Extended Java Dataset.}
\label{fig:extended-java}
\end{figure}

\begin{figure}[!ht]
\centering{\includegraphics[width=\linewidth]{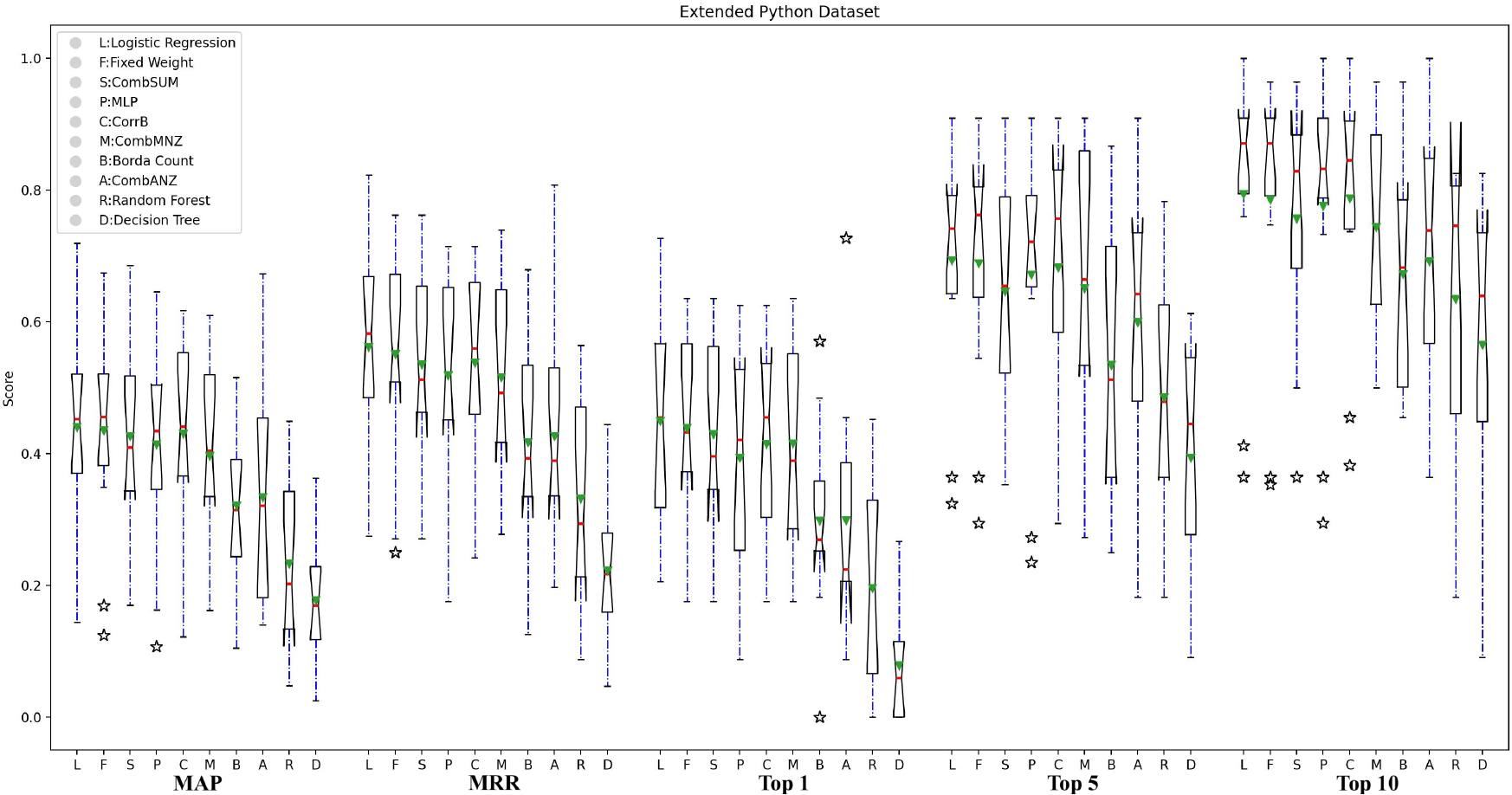}}
\caption{Performance of Different Aggregation Methods on Extended Python Dataset.}
\label{fig:extended-python}
\end{figure}

To assess the performance of different composers in combining three components, we employed both supervised and unsupervised methods, including DT, MLP, RF, LR, fixed weight, CombSUM, CombMNZ, CombANZ, CorrB, and Borda Count, as the composer component. The average results of each composer across the three datasets are presented in Table~\ref{tab:rq3}. The box plots for different composers on the three datasets are depicted in Fig.~\ref{fig:original-java}, Fig.~\ref{fig:extended-java}, and Fig.~\ref{fig:extended-python}, respectively. From the table, we observe that LR, CombSUM, and fixed weight consistently achieve superior results compared to other composers. LR attains the best average outcomes on the original Java dataset and the extended Python dataset. Conversely, DT exhibits the poorest results across all three datasets, with RF following closely behind. \nff{R}esults are even worse than pure TraceScore. \nff{This indicates that the contributions of different component scores are linear, which LR, fixed weight, and CombSUM can model effectively due to their ability to handle linear relationships and lower risk of overfitting. In contrast, RF and DT, being tree-based structures, may struggle to fit the component scores effectively with their decision tree-based splits, which could compromise their performance.}

\begin{figure}[!ht]
\centering
\includegraphics[width=\linewidth]{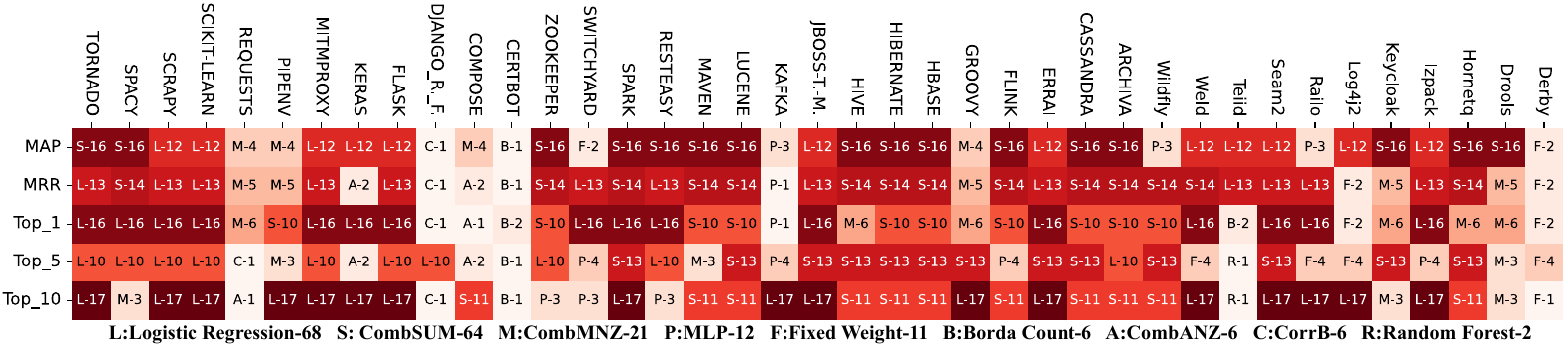}
\caption{Heatmap of the best composer on each project. Each row in the heatmap corresponds to a specific evaluation metric, while each column represents an individual project. For instance, in the first row and first column, ``S-16'' signifies that the ``CombSUM'' composer achieved the best result in the MAP metric for the ``TORNADO'' project. ``16'' means that, across all 39 projects, CombSUM yielded the optimal results in MAP 16 times.
}
\label{fig:heatmap}
\end{figure}

To \nff{visualize} the performance of various composers, we generated the heatmap shown in Fig.~\ref{fig:heatmap}. This heatmap details the top-performing composer for each project across different evaluation metrics. Analysis of the heatmap reveals that, across all projects and the five evaluation metrics, LR demonstrated the highest performance on 68 cases, closely followed by CombSUM with 64 instances. Interestingly, DT never attained the top-performing result. In summary, however, there is no composer that consistently performs best across all projects.

Based on the results above, the question that arises is: for real-world scenarios, which composer to select? The choice of a composer depends on the characteristics of the data, the nature of the retrieval methods, \nff{the relationship between the contributions of different components,} and other desired aspects. Therefore, it is recommended to experiment with different composer methods \nff{(especially those good at handling linear relationships)} and evaluate their performance to determine which one works best in a specific application context.

Across all composers, the average results on the extended Python dataset surpass those on the extended Java dataset, which, in turn, outperform the results on the original dataset. The MAP and MRR on the Python dataset reach as high as 44\% and 56.2\%, respectively. This result suggests that bug localization might be easier on the Python dataset than on the Java dataset. 
Moreover, we observed that Python projects have a smaller size, as measured by the number of source code files, in comparison to Java projects. This observation instigated an exploration into whether a correlation exists between bug localization accuracy and project size. Table~\ref{tab:pearson} presents the Pearson correlation between project size and localization accuracy across all 39 projects. Given that all the p-values are greater than 0.05 (except for MAP of LR), we are unable to reject the null hypothesis. There is insufficient evidence to assert the presence of a significant effect or relationship between localization accuracy and project size.

\begin{table}[htbp]
\setlength{\abovecaptionskip}{0.cm}
\caption{Average Results of Different Composers.}
\label{tab:rq3}
\begin{tabular}{lccccc}
\hline \hline
\multicolumn{6}{c}{Original Java dataset}                  \\ \hline
             & MAP   & MRR   & Top 1 & Top 5 & Top 10 \\
LR           & \textbf{0.302} & \textbf{0.435} & \textbf{0.321} & 0.561 & \textbf{0.650} \\
CombSUM      & 0.300 & 0.433 & 0.321 & \textbf{0.566} & 0.643  \\
Fixed Weight & 0.298 & 0.433 & 0.320 & 0.563 & 0.640  \\
MLP          & 0.298 & 0.430 & 0.316 & 0.560 & 0.642  \\
CorrB        & 0.289 & 0.418 & 0.303 & 0.550 & 0.634  \\
CombMNZ      & 0.277 & 0.416 & 0.314 & 0.527 & 0.627  \\
Borda Count  & 0.219 & 0.356 & 0.262 & 0.464 & 0.543  \\
CombANZ      & 0.195 & 0.292 & 0.191 & 0.387 & 0.479  \\
RF           & 0.179 & 0.279 & 0.161 & 0.394 & 0.526  \\
DT           & 0.056 & 0.077 & 0.023 & 0.090 & 0.171  \\
\hline \hline
\multicolumn{6}{c}{Extended-Java dataset}                  \\ \hline
             & MAP   & MRR   & Top 1 & Top 5 & Top 10 \\ 
CombSUM      & \textbf{0.359} & \textbf{0.498} & \textbf{0.379} & \textbf{0.641} & \textbf{0.729}  \\
Fixed Weight & 0.344 & 0.477 & 0.356 & 0.621 & 0.714  \\
LR           & 0.342 & 0.473 & 0.352 & 0.615 & 0.711 \\
CombMNZ      & 0.330 & 0.473 & 0.353 & 0.609 & 0.700  \\
CorrB        & 0.337 & 0.466 & 0.344 & 0.609 & 0.710  \\
MLP          & 0.300 & 0.417 & 0.304 & 0.541 & 0.641  \\
Borda Count  & 0.257 & 0.396 & 0.293 & 0.515 & 0.590  \\
CombANZ      & 0.247 & 0.353 & 0.240 & 0.474 & 0.589  \\
RF           & 0.203 & 0.295 & 0.170 & 0.428 & 0.564  \\
DT           & 0.064 & 0.083 & 0.015 & 0.121 & 0.225  \\ \hline \hline
\multicolumn{6}{c}{Extended-Python dataset}                  \\ \hline
                   & MAP         & MRR   & Top 1 & Top 5 & Top 10 \\
LR                 & \textbf{0.440}       & \textbf{0.562} & \textbf{0.449} & \textbf{0.693} & \textbf{0.794} \\
Fixed Weight & 0.435 & 0.551 & 0.439 & 0.688 & 0.785\\
CorrB              & 0.430       & 0.538 & 0.414 & 0.682 & 0.787  \\
CombSUM            & 0.426       & 0.535 & 0.429 & 0.646 & 0.756  \\
MLP                & 0.414       & 0.519 & 0.394 & 0.672 & 0.776  \\
CombMNZ            & 0.397       & 0.516 & 0.415 & 0.651 & 0.744  \\
CombANZ            & 0.334       & 0.427 & 0.298 & 0.599 & 0.691  \\
Borda Count        & 0.321       & 0.417 & 0.298 & 0.535 & 0.672  \\
RF                 & 0.233       & 0.332 & 0.196 & 0.485 & 0.635  \\
DT                 & 0.177       & 0.222 & 0.079 & 0.394 & 0.565  \\ \hline \hline
\end{tabular}
\end{table}

\begin{table}[htbp]
\caption{Pearson Correlation Coefficient of Project Size and Localization Accuracy.}
\label{tab:pearson}
\begin{tabular}{ccc|cc|cc}
\hline
\multirow{2}{*}{\begin{tabular}[c]{@{}c@{}}\textbf{Pearson}\\ \textbf{Correlation}\end{tabular}} & \multicolumn{2}{c|}{\textbf{Fixed weight}}      & \multicolumn{2}{c|}{\textbf{LR}}                & \multicolumn{2}{c}{\textbf{CombSUM}}            \\ 
& \textbf{Correlation} & \textbf{p-value} & \textbf{Correlation} & \textbf{p-value} & \textbf{Correlation} & \textbf{p-value} \\ \hline
\textbf{MAP}                                                                            & -0.306                  & 0.058        & -0.364                  & 0.023        & -0.274                  & 0.091        \\
\textbf{MRR}                                                                            & -0.182                  & 0.268        & -0.290                  & 0.074        & -0.100                  & 0.546        \\
\textbf{Top 1 }                                                                         & -0.190                  & 0.246        & -0.309                  & 0.056        & -0.104                  & 0.530        \\
\textbf{Top 5 }                                                                         & -0.158                  & 0.337        & -0.240                  & 0.141        & -0.049                  & 0.767        \\
\textbf{Top 10}                                                                         & -0.174                  & 0.289        & -0.248                  & 0.128        & -0.093                  & 0.572        \\ \hline
\end{tabular}
\end{table}

\answer{
\textit{\textbf{Answering RQ3:} Among all the composers, LR, CombSUM, and fixed weight demonstrate favorable performance across the three datasets, while DT and RF significantly lag behind other composers. Hence, we highlight that the original fixed weights introduced in AmaLgam provide a simple, immediately applicable composition configuration for most settings. Nevertheless, it is advisable to explore various methods in order to select a composer based on specific situations.}}

\subsection{Discussion}

Overall, as shown in Table~\ref{tab:summary}, our experimental results suggest that TraceScore is \textbf{replicable under \textit{relaxed cut-off date} constraint}, but, \textbf{non-replicable under \textit{strict cut-off date} constraint}, where the former can achieve better results. However, in actual applications, the choice between relaxed cut-off date and strict cut-off date is flexible, as commits available at the time when developers perform bug-fixing tasks will be considered for recommendation.

On the extended dataset, TraceScore also yields similar results compared with on the original dataset, which demonstrates that \textbf{TraceScore possesses good generalizability}. However, the results vary more (i.e., some projects exhibit much higher performance, other projects exhibit even lower performance), \nff{which means} it is not possible to accurately predict the performance of TraceScore on a new project. Additional investigations are necessary to understand when TraceScore is expected to perform well and under which conditions TraceScore will not yield a lot of benefits.

\textbf{ABLoTS}, in contrast, \textbf{is not reproducible} for two main reasons: 1) the authors reused the wrong BugCache implementation from Wang et al.~\cite{wang2014version} (we confirmed the incorrect use with Wang et al.), which results in the BugCache score greatly boosting the final result; 2) when we adopt the correct BugCache score, we could not duplicate the DT composer because of a lack of details in the original study. We are skeptical whether DT is the right choice for the composer, as also different sampling strategies and hyperparameter tuning yielded a performance worse than the static composer configuration. When we utilized this fixed weight composer proposed by AmaLgam we observed its performance to hold also for the extended dataset.

Both TraceScore and the fixed weight-based ABLoTS demonstrate superior results on the extended Python dataset compared to the other two Java datasets. What is particularly noteworthy is that, despite the absence of feature requests and traceability information in the Python dataset, TraceScore still achieves better localization results than on the other two Java datasets. A test examining the correlation between project size and bug localization performance indicated that such a correlation does not exist (refer to Table.~\ref{tab:pearson} for the Pearson correlation coefficient of project size and bug localization performance). Hence we believe that further investigation into bug reports and bug locations is required to understand which factors, such as textual alignment of terms used in issues and source code, or cohesion of terms, or diversity of terms help to explain bug localization performance.

We observed that combining all three scores can improve the TraceScore result by 50\% $\sim$ 105\% on both datasets, which suggests that a combination of different components is likely to outperform any single mechanism. To this end, the choice of composer is a crucial aspect. Our empirical evaluation of the performance of various composers unveiled that LR, CombSUM, and fixed weight consistently outperformed, while DT and RF exhibited subpar performance.

One additional take-away of our replication study is \nff{to pay} attention to the challenge of properly evaluating a technique in the presence of temporal aspects, especially when third-party research outcomes (i.e., BugCache) behave differently than expected. The case of the 15-day interval of BugCache is especially tricky, as for other datasets where commits of a bug predominately happen more than 15 days before the bug's closing date, no such negative side effect would have been noticeable.
In the case of ABLoTS sanity checks on the DT's feature importance values would have identified unexpected results (i.e., with BugCache rather than TraceScore dominating the classification result), subsequently triggering confirmation or revision of the composer mechanism. 

Overall, the results of this replication study suggest that the state of the art in bug localization is not as useful as prior results have suggested and that further research is still needed to obtain results that are good enough to be useful in practice.

\begin{table}[htbp]
\setlength{\abovecaptionskip}{0.cm}
\caption{Summary of Results.}
\label{tab:summary}
\begin{tabular}{|cl|c|c|}
\hline
\multicolumn{2}{|c|}{}                                                   & Replicable & Generalizable \\ \hline
\multicolumn{1}{|c|}{\multirow{2}{*}{TraceScore}} & Relaxed cut-off date & Yes        & Yes           \\ 
\multicolumn{1}{|c|}{}                            & Strict cut-off date  & No         & -             \\ \hline
\multicolumn{2}{|c|}{ABLoTS}                                             & No         & -             \\ \hline
\multicolumn{2}{|c|}{Fixed Weight Composer}                              & -          & Yes           \\ \hline
\end{tabular}
\end{table}

\subsection{Implications to Future Replication }

In this section, we summarize lessons learned through our replication specific to bug localization. The goal is to support researchers in more efficiently and rapidly replicating the approach for a comparative study or as a baseline for novel approaches. 

\begin{itemize}
    \item \textbf{Data Collection:} During the data collection, Rath et al. collected both bug reports and non-bug reports, traceability information between reports, commit logs, commit code change, and constructed links from issues to code change. For the ground truth construction, Rath et al. utilized the modified files and newly added files as the ground truth. However, in our opinion, which files are newly added cannot be predicted by definition. In contrast, removed files are predictable, and should be included in the ground truth. \nff{This slight difference in the construction of the ground truth would not change the approach, but could potentially affect the evaluation results.}
    \item \textbf{Trace Graph Construction:} The construction of the trace graph requires previously fixed issues. When replicating, researchers need to keep in mind that bug fixing (or development efforts in the scope of an issue in general) are not atomic events but potentially long-lasting activities with days, sometimes even months, between issue creation date, commit dates, and issue closing dates.  Hence, they need to be careful about the date for artifacts selection. \nff{When predicting, only information available at that time should be used.}
    
    \item \textbf{BugCache Calculation:} For selecting the historical bug-fixing commits, apart from keywords-based selection, also a bug ID can identify bug-fixing commits. More important, as with the trace graph construction, any commit information taken for file candidate scoring must have been already available by the fictive recommendation time (e.g., bug creation date) or at the latest the bug's first partial fix implementation. As we have seen in the replication study, using the bugs' fixed date may lead to data leakage.
    \item \textbf{Choice of Composer:} Given a limited set of features (i.e., the three suspiciousness scores), a DT may not be the most suitable option. Researchers could initiate the exploration with LR, CombSUM, and the fixed weight composer. However, it is crucial to experiment with various methods to select a composer tailored to specific situations.
\end{itemize}

\section{Threats to Validity}\label{sec:validity}

In this section, we discuss the main threats to the validity of our results, and how we mitigate them. We use the taxonomy of Wohlin et al.~\cite{Wohlin2000}.

\textbf{Construct Validity.} One possible threat to construct validity is that there is no available open source implementation of ABLoTS approach, which means we had to re-implement it by ourselves. To alleviate this, we carefully read the original study, trying to reproduce it as close as possible. For the BLUiR component, we reused existing open source code from a published paper to reduce possible errors. As for BugCache component, we translate the original implementation from Java into Python with great care. We carefully examined the code and the output to avoid errors. \nff{Furthermore, we reached out to the authors of both ABLoTS and AmaLgam approaches to clarify experiment details, such as the selection of the cut-off date, aiming to ensure precise replication.}
Another potential construct validity concern arises from the dataset. The Python projects are sourced from the BuGL~\cite{muvva2020bugl} dataset, a large-scale cross-language dataset for bug localization. The Python dataset lacks feature requests and traceability information, potentially introducing bias to the performance evaluation of TraceScore.

\textbf{Internal Validity.} From a perspective of internal validity, potential errors can happen in the reproduction (e.g., settings and library usage), which is a common threat to replication studies. We tried out possible settings and compare\nff{d} the results with the original study. Another potential threat is that the open source projects in our dataset might have been changed by the day we collected from GitHub. To address this threat, we filter out projects that do not have complete commit information anymore. 

\textbf{External Validity.} Regarding external validity, we experimented only on open source Java projects and Python projects. We encourage future studies to replicate this study with commercial projects to be able to obtain insights into ABLoTS' generalizability in industrial settings.

\textbf{Conclusion Validity.} Conclusion Validity could come from the interpretation of the results, which includes the evaluation metrics for evaluation and K-S test for comparison. To mitigate the threat, we adopted the same evaluation metrics adopted in the original paper. Then the two sample K-S test was utilized to compare the difference of experiment results, as it is sensitive to differences in both location and shape of the empirical cumulative distribution functions of the two samples.

\section{Related Work}\label{sec:relatedwork}
Several IRBF approaches have been proposed in the last years, which leverage information retrieval techniques to find buggy-prone snippets from all source code candidates. We describe below the approaches related to our work.

BugLocator calculates similarity between bug reports to recommend similar files to similar bug reports~\cite{zhou2012should}. Sisman and Kak propose a source code version history-based fault localization approach, which utilizes the frequency of a file being buggy and its modifications to prioritize candidate source code files~\cite{sisman2012incorporating}. Wang et al. combine similar bug reports, code version history, and code structure to find the buggy files~\cite{wang2016amalgam+, wang2014version}. 
Lucia et al. investigate five data fusion methods to improve spectrum-based fault localization techniques~\cite{lucia2014fusion}. 

Niu et al. propose a refactoring-aware traceability model for constructing more accurate code history, which can boost the results of similar bug reports and code version history component~\cite{niu2023rat}. Wen et al. use change logs and change hunks from commit message as alternative of segments of source code files to enable more accurate bug localization~\cite{wen2016locus}. 

\nff{Recently, with the emergency of deep learning (DL),
many IRBL approaches utilizing DL techniques have been proposed~\cite{ciborowska2022fastfbl-bert, han2023bjxnet,yong2023decomposings-buglocator}. They compute the semantic similarity between source code and bug reports by converting them into deep representations. They achieve better results than methods relying solely on code structure. However, similar reports component and version history component continue to play indispensable roles. Many approaches still combine results from the similar reports component~\cite{lam2017bugdnnloc, xiao2019improvingdeeploc, sangle2020drast, cao2020bugpecker, yang2021locatingmram, qi2021dreamloc, shi2022semirfl, xu2023buglocfront} and version history component~\cite{lam2015combining, lam2017bugdnnloc, xiao2019improvingdeeploc, sangle2020drast, wang2020multimd-cnn, cao2020bugpecker, anh2021imbalanced, yang2021locatingmram, qi2021dreamloc,shi2022semirfl, xiao2023bugradar, xu2023buglocfront, Al-Aidaroosimpact} with DL-based semantic similarity to achieve improved bug localization effectiveness. For example, the MD-CNN approach~\cite{wang2020multimd-cnn} combines similar bug reports and bug-fixing history with other code structure similarities, employing convolutional neural networks to extract features for IRBL.}

For comparison of state-of-the-art approaches, Garnier and Garcia evaluate the effectiveness of BLUiR~\cite{saha2013improving}, BLUiR+~\cite{saha2013improving}, and AmaLgam~\cite{wang2014version} on 20 C\# projects~\cite{garnier2016evaluation}. Lee et al. conducted a generalized and large-scale investigation into six IRBL techniques~\cite{bench4bl}.
Akbar et al. divided IRBL tools into three generations and presented a comprehensive large-scale evaluation of all three generations of bug-localization tools with code libraries in multiple languages (including Java, C, C++ and Python)~\cite{akbar2020large}. Li et al. re-implement six state-of-the-art bug localization approaches and report their effectiveness on 10 Huawei projects~\cite{li2022empirical}. Lee et al.~\cite{bench4bl} and Li et al.~\cite{li2022empirical} analyzed the same five state-of-the-art approaches and found lower average results than the original ABLoTS results (e.g., MAP less than 0.4, and MRR less than 0.53).

Despite these several empirical studies described above, none of them included ABLoTS, which strengthens the usefulness of our study. Moreover, to the best of our knowledge, there is no study that investigate the performance of different composers on combining the suspiciousness scores of the three components. Thus, our work address both limitation of related work, by replicating ABLoTS and exploring the use of different composers.


\section{Conclusion}\label{sec:conclusion}

In this paper, we conduct a replication study of the ABLoTS approach for bug localization. We recreated the original approach, both on the original dataset and two extended datasets (one Java dataset and one Python dataset). Furthermore, we also conduct empirical analysis on the performance of various composers. We found that the core component of ABLoTS, i.e., TraceScore, is replicable and generalizable under a \textit{relaxed cut-off} constraint, but irreplicable under a \textit{strict cut-off} constraint. ABLoTS is neither replicable nor generalizable because of the adoption of an incorrect cut-off date in the BugCache subcomponent, leading to test data leaking into training data. Also, the chosen technique to combine multiple scores yielded poor results when applied to the correctly derived scores. Our study emphasizes the importance of choosing the proper cut-off dates in evaluating bug localization techniques. The comparison between different programming languages suggests that bug localization on Python projects might be more straightforward than on Java projects. Our empirical analysis reveals that LR, fixed weight, and CombSUM demonstrate superior performance in combining different components, while DT and RF exhibit poor performance.

As part of future work, we will start investigating alternative information sources and techniques to improve bug localization performance. We will also explore the effectiveness of large language models in bug localization, aiming to achieve higher accuracy in bug localization.

\bmhead{Data Availability}
\label{sec:data}
Artifacts of the experiment are available in an online replication package: \href{https://github.com/feifeiniu-se/Replication2}{https://github.com/feifeiniu-se/Replication2}



\bmhead{Acknowledgments}

This work is supported by Natural Science Foundation of Jiangsu Province, China (BK20201250), Cooperation Fund of Huawei-NJU Creative Laboratory for the Next Programming, and also supported in part by NSF Grant 2034508 (USA), by a Sam Taylor Fellowship Award,  the Austrian Science Fund (FWF) grant P31989-N31 and P34805-N as well as the LIT Secure and Correct System Lab sponsored by the province of Upper Austria.

\section*{Compliance with Ethical Standards}

\textbf{Conflict of Interest}
The authors declared that they have no conflict of interest.

\clearpage
\section*{Author Biography}

\begin{biography}{\includegraphics[width=0.14\textwidth]{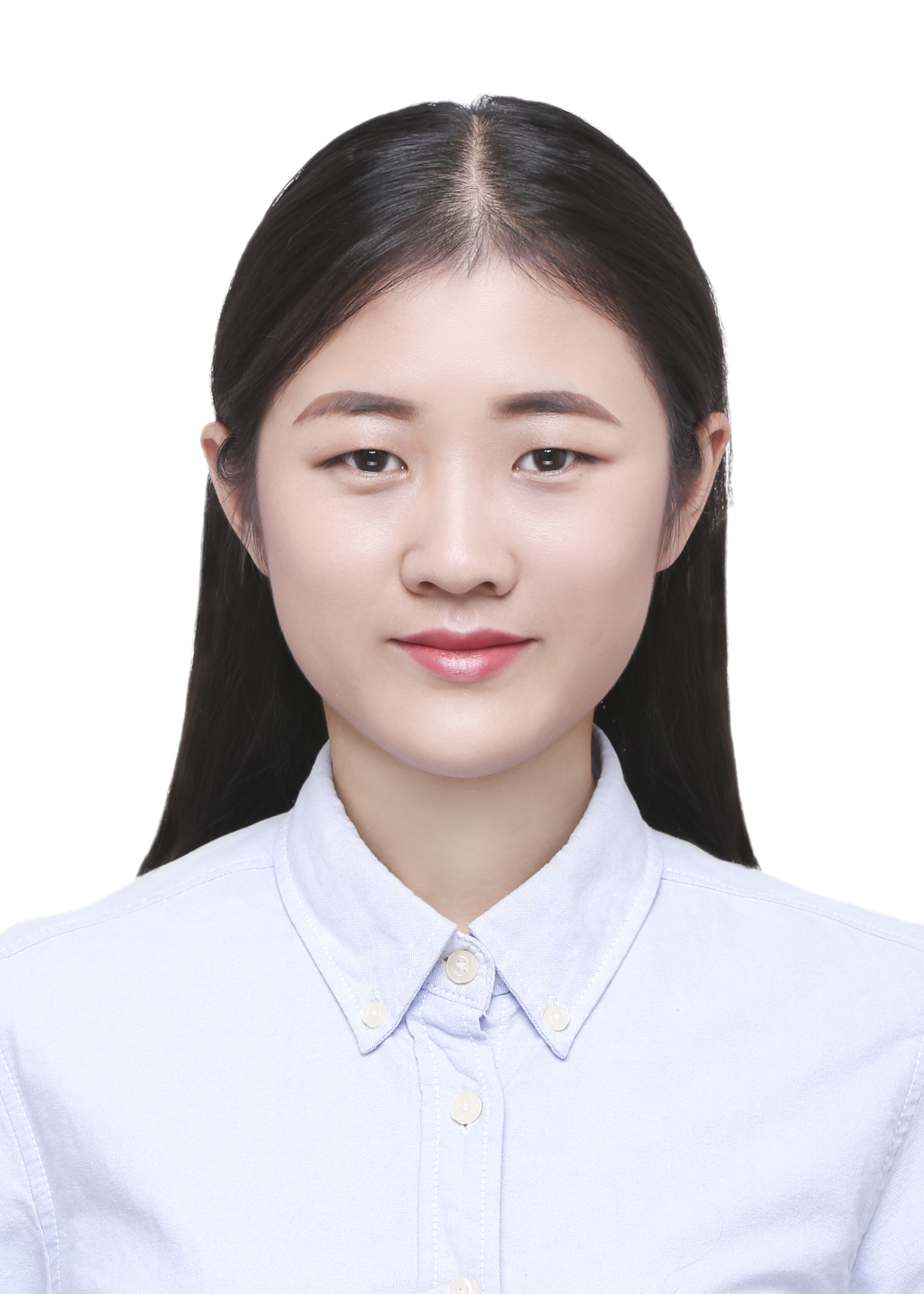}}{\textbf{Feifei Niu} is a Research Fellow at the University of Ottawa. She received her Doctorate from Nanjing University. Her research interests include software quality assurance, software testing, requirements engineering.}
\end{biography}

\vspace{10mm}

\begin{biography}{\includegraphics[width=0.14\textwidth]{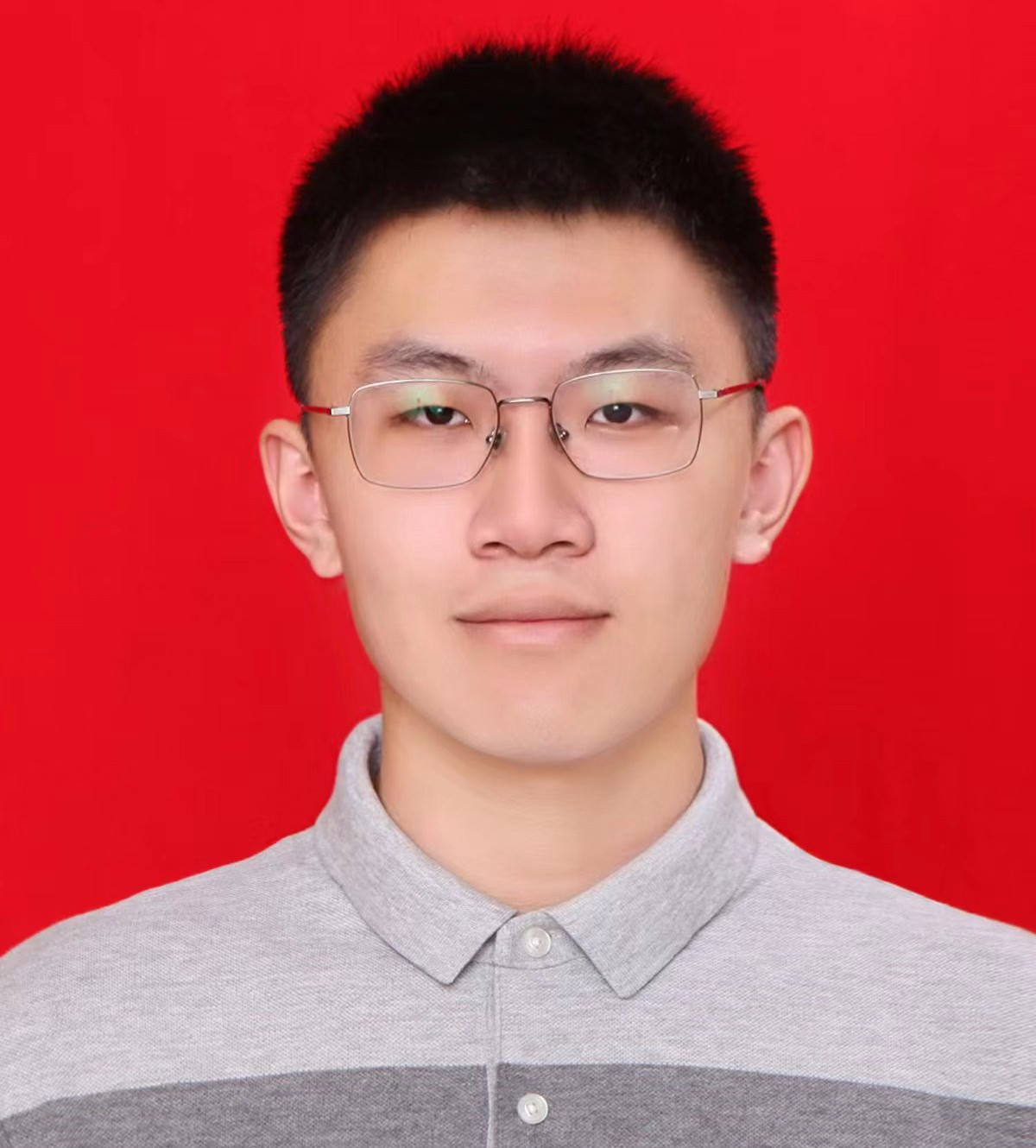}}{\textbf{Enshuo Zhang} is a master student at Nanjing University. His research interests include software engineering, natural language processing. He can be contacted at 2575357413@qq.}
\end{biography}

\vspace{5mm}

\begin{biography}{\includegraphics[width=0.14\textwidth]{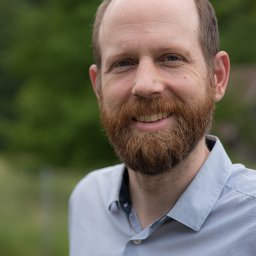}}{\textbf{Christoph Mayr-Dorn} is a senior researcher at the Institute for Software Systems Engineering at the Johannes Kepler University Linz, Austria. He holds a Ph.D. in Computer Science from the Technical University Vienna. His current research interests include software process monitoring and mining, change impact assessment, and software engineering of cyber–physical production systems.}
\end{biography}

\begin{biography}{\includegraphics[width=0.14\textwidth]{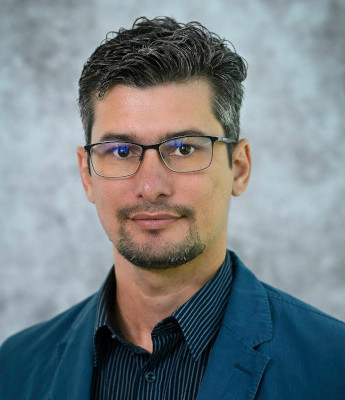}}{\textbf{Wesley Klewerton Guez Assunção}
is an Associate Professor with the Department of Computer Science at North Carolina State University. Wesley was a University Assistant in the Institute of Software Systems Engineering (ISSE) at Johannes Kepler University Linz (JKU), Austria (2021–2023); a Postdoctoral Researcher at Pontifical Catholic University of Rio de Janeiro (PUC-Rio), Brazil (2019–2023); and an Associate Professor at Federal University of Technology - Paraná, Brazil (2013 to 2020). He obtained his M.Sc. and Ph.D. in Computer Science from Federal University of Paraná (UFPR) also in Brazil. 
Further information: https://wesleyklewerton.github.io/.}
\end{biography}

\begin{biography}{\includegraphics[width=0.14\textwidth]{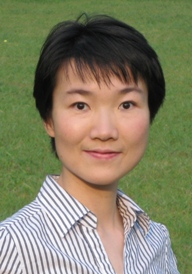}}{\textbf{Liguo Huang} received the MS and PhD degrees from the Computer Science Department and Center for Systems and Software Engineering (CSSE), the University of Southern California (USC). She is an Associate Professor with Computer Science Department, the Southern Methodist University (SMU), Dallas, TX, USA. Her primary research centers around synergy of machine/deep learning, natural language processing and software engineering, software quality assurance, process modeling and improvement, stakeholder/value-based software engineering.}
\end{biography}

\begin{biography}{\includegraphics[width=0.14\textwidth]{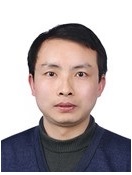}}{\textbf{Jidong Ge} is an Associate Professor at Software Institute, Nanjing University. He received his PhD degree in Computer Science from Nanjing University in 2007. His current research interests include workflow modeling, process mining, cloud computing, workflow scheduling, software engineering. His research results have been published in more than 90papers in international journals and conference proceedings including IEEE TPDS, IEEE TSC, JASE, COMNET, JPDC, FGCS, JSS, Inf. Sci., JNCA, JSEP, ESA, ICSE, IWQoS etc}
\end{biography}
\vspace{0mm}

\begin{biography}{\includegraphics[width=0.14\textwidth]{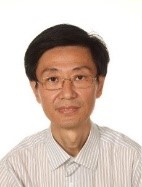}}{\textbf{Bin Luo} is a Full Professor at the Software Institute, Nanjing University. His main research interests include cloud computing, computer network, workflow scheduling, and software engineering. His research results have been published in more than 50 papers in international journals and conference proceedings including IEEE TSC, ACM TIST, JSS, FGCS, Inf Sci, ESA, ICTAI, etc. He is leading the institute of applied software engineering at Nanjing University.}
\end{biography}
\vspace{0mm}

\begin{biography}{\includegraphics[width=0.14\textwidth]{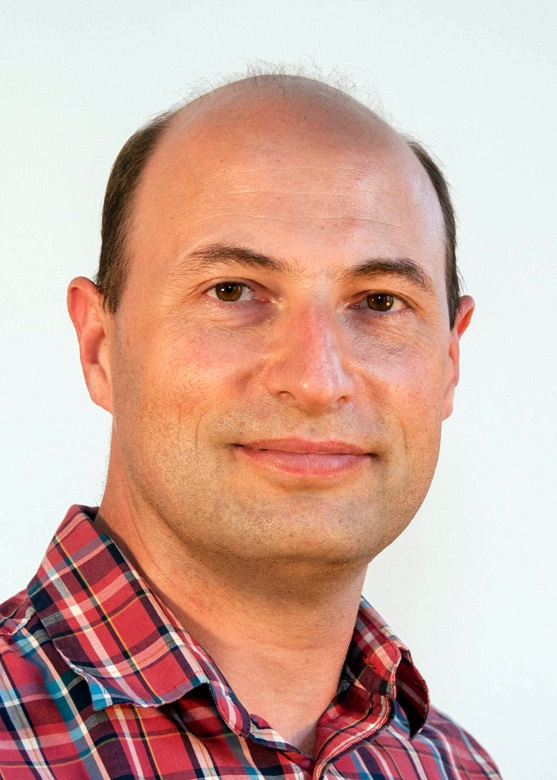}}{\textbf{Alexander Egyed} is a Full Professor and Chair for Software-Intensive Systems at the Johannes Kepler University, Austria (JKU). He received a Doctorate degree from the University of Southern California, USA in 2000 and then worked for industry for many years before joining the University College London, UK in 2007 and JKU in 2008. He is most recognized for his work on software and systems design — particularly on variability, consistency, and traceability.}
\end{biography}

\end{document}